# Ensemble Properties of RVQ-Based Limited-Feedback Beamforming Codebooks

Vasanthan Raghavan* and Venugopal V. Veeravalli

**Abstract**

The ensemble properties of *Random Vector Quantization (RVQ)* codebooks for limited-feedback beamforming in multi-input multi-output (MIMO) systems are studied with the metrics of interest being the received SNR loss and mutual information loss, both relative to a perfect channel state information (CSI) benchmark. The simplest case of unskewed codebooks is studied in the correlated MIMO setting and these loss metrics are computed as a function of the number of bits of feedback ($B$), transmit antenna dimension ($N_t$), and spatial correlation. In particular, it is established that: i) the loss metrics are a product of two components – a quantization component and a channel-dependent component; ii) the quantization component, which is also common to analysis of channels with independent and identically distributed (i.i.d.) fading, decays as $B$ increases at the rate $2^{-B/(N_t-1)}$; iii) the channel-dependent component reflects the condition number of the channel. Further, the precise connection between the received SNR loss and the squared singular values of the channel is shown to be a Schur-convex majorization relationship. Finally, the ensemble properties of skewed codebooks that are generated by skewing RVQ codebooks with an appropriately designed fixed skewing matrix are studied. Based on an estimate of the loss expression for skewed codebooks, it is established that the optimal skewing matrix is critically dependent on the condition numbers of the *effective channel* (product of the true channel and the skewing matrix) and the skewing matrix.

V. Raghavan is currently with the University of Southern California, Los Angeles, CA 90089, USA. He was with the Coordinated Science Laboratory, University of Illinois at Urbana-Champaign, Urbana, IL 61801, USA when this work was done. V. V. Veeravalli is with the Coordinated Science Laboratory and the Department of Electrical and Computer Engineering, University of Illinois at Urbana-Champaign, Urbana, IL 61801, USA. Email: vasanthan_raghavan@ieee.org, vvv@illinois.edu. *Corresponding author.

This work was supported in part by the National Science Foundation through grant CNS-0831670. This paper was presented in part at the IEEE International Symposium on Information Theory, Seoul, South Korea, 2009.

# I. Introduction

Optimal signalling to maximize the achievable rate in multi-input multi-output (MIMO) communication channels requires appropriate adaptation of the number of transmit data-streams in response to the SNR, channel correlation, and the channel state information (CSI) available at the transmitter and the receiver [1], [2]. On the other hand, an increase in the number of transmit data-streams results in a significant increase in the number of radio-frequency (RF) link chains and imposes a corresponding increase in complexity and cost [3]. Thus, in many later generation (3G/4G and beyond) cellular standards such as WiMAX, 3GPP-LTE, etc., low-complexity signalling alternatives are preferred. In particular, *beamforming*, where the number of transmit data-streams is fixed to be one (*independent* of the SNR, channel correlation or CSI) is an attractive choice due to its low-complexity. Beamforming is also preferred when the central goal is to maximize the coverage area/range of signalling, over the $60$ GHz regime [4] where a large number of small antennas can be packed in a fixed area to reap the array gain possible with beamforming, and as a mechanism for cross-layer signalling in ad-hoc networks.

***Background:*** The performance achieved with a beamforming scheme is clearly dependent on the quality of CSI available at both the transmitter and the receiver. While perfect CSI at the receiver is a reasonable assumption for practical systems, constraints on channel tracking and quality of feedback ensure that perfect CSI at the transmitter is an optimistic assumption. Nevertheless, the possibility of low-rate reverse link feedback from the receiver to the transmitter has resulted in the popularity of *limited-feedback* systems [5], [6], where $B$ bits of channel quality information are fed back to the transmitter. The common method of using the feedback resource in beamforming systems is by designing a codebook of $2^B$ beamforming vectors and feeding back the index of the best codeword from the codebook over each coherence period [5], [6].

Given a channel correlation profile, the problem of optimal design of $B$-bit codebooks is ill-posed (in general) and hence, difficult. In the special case of channels with independent and identically distributed (i.i.d.) fading, Grassmannian constructions that are designed to maximize the minimum distance between beamforming vectors have been proposed in [7] and [8]. The intuition behind this proposal is that the dominant right singular vector of an i.i.d. channel is isotropically (uniformly) distributed in the space of $N_t$-dimensional unit-norm beamforming vectors where $N_t$ is the number of transmit antennas. Thus, a "good" limited-feedback codebook is an efficient quantization of this ambient space. Grassmannian codebooks are obtained via algebraic techniques [9]–[11] and are technically impossible to construct for some $(N_t, B)$-combinations.

To overcome this difficulty, inspired by the random coding argument, *Random Vector Quantization* (RVQ) codebooks have also been proposed in the literature [12]. RVQ codebooks were first introduced in the context of signature matrix quantization for Code-Division Multiple Access (CDMA) systems in [13], [14]. RVQ codebooks are instantiations of random constructions (in contrast to Grassmannian codebooks) and the beamforming vectors are isotropic and i.i.d. over the ambient space. Thus, RVQ codebooks can be designed for all $(N_t, B)$-combinations and they are of low-complexity in terms of design. The intuition behind an RVQ codebook design has been extended to the multi-user setting (with i.i.d. fading) in many recent papers [15]–[19].

In the general single-user setting where the channel matrix is spatially correlated and the dominant right singular vector of the channel has certain preferred directions, Grassmannian codebooks are *mismatched* and are hence, sub-optimal. In fact, in [20, Figs. 6 and 7], [21] illustrative examples are given, where Grassmannian codebooks suffer dramatic performance losses (on the order of $25$ dB in $\mathsf{SNR}$) relative to the perfect CSI benchmark. In these situations, more complicated (in terms of design) *spherical* Vector Quantization (VQ) constructions [22]–[24] based on the Lloyd algorithm have been proposed. While VQ codebooks are optimal[1], it is hard to obtain insights on the structure of the optimal codebook. To overcome these difficulties, rotation and scaling-based codebooks have been proposed [20], [25]–[29] and shown to result in significant improvement in performance over Grassmannian codebooks. The main idea behind these constructions is to finely quantize the local neighborhood around the statistically dominant eigen-directions and coarsely quantize elsewhere (if $B$ is large enough to afford this possibility).

Towards the eventual goal of an optimal codebook construction, it is imperative to understand the performance of existing codebook designs and identify the merits/demerits of existing schemes with respect to fundamental limits on performance. In this direction, the performance of an ensemble of RVQ codebooks has been studied for i) i.i.d. multi-input single-output (MISO) channels [12], [15], [30], ii) correlated MISO channels in the asymptotic-$B$ regime via high resolution quantization theory [31], [32], iii) i.i.d. MIMO channels via bounds [33], [34], iv) i.i.d. MISO and MIMO channels in the large antenna regime via extreme order statistics [12], [35], and v) symbol error rate of limited-feedback beamforming in an i.i.d. MISO setting [36], [37].

Both exact expressions as well as asymptotic approximations (in $B$) are available for RVQ codebooks for MISO channels in both the i.i.d. and correlated settings and these studies show that the rate of decay of the loss metrics is of the order of $2^{-\frac{B}{N_t-1}}$ as $B$ increases. However, in the MIMO setting, performance analysis is available only in the i.i.d. case in the large antenna regime. Further, since reverse link feedback is a valuable resource, the practically relevant regime is when $B$ is small and there has been little to no attention in the literature on performance analysis relevant to this regime. More importantly, to the best of our knowledge, the performance of non-RVQ codebooks has not been studied at all. Thus, it is of interest to understand the ensemble properties of RVQ codebooks (as well as codebooks designed based on RVQ codebooks and tailored for correlated channels) in the most general correlated setting for practically relevant values of $B$.

[1]Technically, VQ codebooks meet the necessary conditions for an optimal codebook construction, but not the sufficient condition. Nevertheless, it is widely believed that VQ constructions are optimal.

***Contributions:*** The main goal of this work is to study the performance of a $B$-bit RVQ codebook in correlated MIMO channels with the metrics of interest being the received $\mathsf{SNR}$ loss ($\Delta\mathsf{SNR}_{\mathsf{rx}}$) and loss in average mutual information ($\Delta I$), both relative to a perfect CSI scheme. For this, we adopt a program of first averaging the loss metric (with a fixed channel realization) over the randomness in the RVQ codebook structure and then, averaging over the randomness in the channel. In this direction, we identify the structure of the density function of the weighted-norm of isotropically distributed unit-norm vectors. With this information, we obtain closed-form expressions (although the results are modulo averaging over channel randomness) for $\Delta\mathsf{SNR}_{\mathsf{rx}}$ and $\Delta I$. The fundamental contributions of this work are three-fold: i) the loss expressions are accurate for small values of $B$ across a large family of channels, ii) they are asymptotically tight in $B$ and the rate of decay with $B$ is still $2^{-\frac{B}{N_t-1}}$ in correlated MIMO channels, and iii) they capture the impact of the channel correlation structure on the performance of RVQ codebooks.

Further, we also establish a *continuous* mapping from the space of all majorizable channels to performance loss with the RVQ codebook in that channel by showing that $\Delta\mathsf{SNR}_{\mathsf{rx}}$ is a Schur-convex function of the squared singular values of the channel. An important consequence of this result is that a channel that is well-conditioned leads to the smallest value for $\Delta\mathsf{SNR}_{\mathsf{rx}}$, whereas a $\mathsf{rank}$-1 channel leads to the largest value for $\Delta\mathsf{SNR}_{\mathsf{rx}}$. As the rank of the channel decreases and/or the condition number of the non-trivial singular values of the channel increases, performance loss with the RVQ codebook relative to a perfect CSI scheme increases. Intuitively, RVQ codebooks are isotropic constructions whereas perfect CSI beamforming corresponds to skewing the signal along the dominant right singular vector of the channel. Thus, a channel that has an isotropically distributed dominant right singular vector (an i.i.d. channel) is best *matched* for the RVQ codebooks, whereas a channel that has a fixed direction for the dominant right singular vector (a $\mathsf{rank}$-1 channel) is poorly *matched* for RVQ codebooks. This intuition mirrors the *source-channel matching* principle for statistical semiunitary precoding established in one of our prior works [21]. Since majorization only results in a partial ordering on the family of all channels, we show that a simplified ordering metric to approximately order and compare the performance of the RVQ scheme (in all channels) is the dominant squared singular value of the channel.

Recent interest in the limited-feedback literature [25], [26] has been on the design of skewed codebooks where a fixed skewing matrix is used to skew an RVQ codebook (or a Grassmannian codebook). The skewing matrix biases the isotropic beamforming vectors in the RVQ codebook and orients them along its singular vectors. Thus, by a suitable choice of the skewing matrix, significant performance improvement can be achieved relative to the RVQ scheme. Despite these observations, technical challenges have ensured that the performance analysis of skewed codebooks has not been addressed in the literature. In the last part of this paper, we overcome this challenge to generalize our characterization of the ensemble properties of RVQ codebooks to the case of skewed codebooks. Our result captures the received $\mathsf{SNR}$ loss in terms of the skewing matrix thus allowing us to obtain insights into the structure of the optimal skewing matrix for limited-feedback beamforming. Our study establishes the criticality of the condition numbers of the *effective channel* (which is the product of the true channel matrix and the skewing matrix) and the skewing matrix in this question. Building on this insight, we construct a class of skewed codebooks that match the left singular vectors of the skewing matrix with the dominant eigen-directions of the transmit covariance matrix of the channel. Numerical studies show that these skewed codebooks significantly out-perform RVQ codebooks and are better than the codebooks proposed in [25], [26].

***Organization:*** This paper is organized as follows. In Section II, we introduce the limited-feedback beamforming setup. In Section III, we study the received SNR loss with an ensemble of RVQ codebooks in the most general (correlated MIMO) setting, whereas in Section IV, our focus is on ordering (comparing) channels with respect to the received SNR loss metric. For this, a partial ordering in the form of a majorization result and an approximate complete ordering are presented in Sec. IV. In Section V, we study the mutual information loss with RVQ codebooks, while in Section VI, we extend the analysis of Sec. III to the skewed codebook setting. Concluding remarks are provided in Section VII. Proofs of most of the results are relegated to the Appendices.

***Notations:*** Upper- and lower-case bold symbols are used to denote matrices and vectors, respectively. The $i$-th element of a vector $\mathbf{x}$ is denoted by $\mathbf{x}(i)$ and its two-norm is denoted as $\|\cdot\|$. The Hermitian transpose of a matrix is denoted by $(\cdot)^\dagger$ while the trace and rank operators are denoted by $\mathsf{Tr}(\cdot)$ and $\mathsf{rank}(\cdot)$, respectively. The eigenvalues of an $N_t \times N_t$ positive semi-definite matrix $\mathsf{M}$ are arranged in decreasing order as $\lambda_1(\mathsf{M}) \geq \cdots \geq \lambda_{N_t}(\mathsf{M})$. Many times, we will find it convenient to write the above relationship as $\lambda_1 \geq \cdots \geq \lambda_{N_t}$ when there is no ambiguity about the matrix under consideration. If $\mathsf{M}$ is a full-rank matrix, the squared condition number $\chi_{\mathsf{M}}$ is defined as $\frac{\lambda_1(\mathsf{M}\mathsf{M}^\dagger)}{\lambda_{N_t}(\mathsf{M}\mathsf{M}^\dagger)}$. We loosely say that $\mathsf{M}$ is ill-(or well-)conditioned depending on whether $\chi_{\mathsf{M}}$ is (or is not) significantly larger than 1. The indicator function and probability of an event are denoted by $\mathbb{1}(\cdot)$ and $\Pr(\cdot)$ while the expectation operator is denoted as $\mathbf{E}[\cdot]$. The symbols $\mathcal{C}$, $B$, $\mathsf{C}_\bullet$, $\mathbf{I}$ and $\mathsf{diag}(\cdot)$ are reserved for limited-feedback codebooks, number of bits of feedback, constants in theoretical statements/results, identity matrix, and a diagonal matrix, respectively. The symbols $\mathbb{C}$ and $\mathbb{R}$ stand for the complex and real fields while $\mathbb{R}^+_n$ and $\mathbb{R}^+$ stand for positive real fields of $n$ and 1 dimensions, respectively. The notations $f(B) \stackrel{B\to\infty}{\asymp} g(B)$ and the little-oh notation $f(B) = o(g(B))$ as $B \to \infty$ stand for $\lim_{B\to\infty} \frac{f(B)}{g(B)} = 1$ and $\lim_{B\to\infty} \frac{f(B)}{g(B)} = 0$.

## II. Beamforming Setup

We consider a communication system with $N_t$ transmit and $N_r$ receive antennas where one data-stream is used for signalling. The baseband model is given by

$$\mathbf{y} = \sqrt{\rho}\,\mathbf{H}\,\mathbf{f}\,s + \mathbf{n} \tag{1}$$

where $\rho$ is the transmit power constraint, the complex Gaussian input $s$ is i.i.d. with zero mean and unit-energy, $\mathbf{H}$ is the $N_r \times N_t$-dimensional channel matrix, and $\mathbf{n}$ is the $N_r$-dimensional proper complex additive white Gaussian noise. In (1), $\mathbf{f}$ is a vector on the complex Grassmann manifold $\mathcal{G}(N_t, 1)$. That is, $\mathbf{f}$ is a $N_t \times 1$ unit-norm vector representing the equivalence class $\{\mathbf{f}e^{j\theta},\ \theta \in [0, 2\pi)\}$.

The main emphasis in this work is on the impact of the channel matrix on limited-feedback performance. For this, we assume that the channel evolves according to a block fading, narrowband model. We further assume a Rayleigh fading (zero mean complex Gaussian) model for the channel coefficients. The second-order statistics are described via a general, mathematically tractable decomposition of the channel [38]:

$$\mathbf{H} = \mathbf{U}_r \, \mathbf{H}_{\mathsf{ind}} \, \mathbf{U}_t^{\dagger} \tag{2}$$

where $\mathbf{H}_{\mathsf{ind}}$ has independent, but not necessarily identically distributed entries, and $\mathbf{U}_t$ and $\mathbf{U}_r$ are unitary matrices that serve as eigen-bases for the transmit and the receive covariance matrices ($\mathbf{\Sigma}_t$ and $\mathbf{\Sigma}_r$), respectively. The covariance matrices are defined as

$$\mathbf{\Sigma}_t \triangleq \mathbf{E}\left[\mathbf{H}^{\dagger}\mathbf{H}\right] = \mathbf{U}_t \, \mathbf{E}\left[\mathbf{H}_{\mathsf{ind}}^{\dagger}\mathbf{H}_{\mathsf{ind}}\right] \mathbf{U}_t^{\dagger} \tag{3}$$

$$\mathbf{\Sigma}_r \triangleq \mathbf{E}\left[\mathbf{H}\mathbf{H}^{\dagger}\right] = \mathbf{U}_r \, \mathbf{E}\left[\mathbf{H}_{\mathsf{ind}}\mathbf{H}_{\mathsf{ind}}^{\dagger}\right] \mathbf{U}_r^{\dagger}. \tag{4}$$

The well-known Kronecker-product correlation model (where $\mathbf{H}_{\mathsf{ind}} = \mathbf{\Lambda}_r^{1/2} \, \mathbf{H}_{\mathsf{iid}} \, \mathbf{\Lambda}_t^{1/2}$ with $\mathbf{H}_{\mathsf{iid}}$ denoting an i.i.d. channel matrix) and virtual representation (where $\mathbf{U}_t$ and $\mathbf{U}_r$ are Fourier matrices) are special cases of (2). Readers are referred to [38], [39] for a detailed study of channel modeling issues.

We study the coherent case with perfect CSI at the receiver. With beamforming, both ergodic capacity and (uncoded) error probability are captured by the received $\mathsf{SNR}$, defined as,

$$\mathsf{SNR}_{\mathsf{rx}} \triangleq \rho \cdot \mathbf{f}^{\dagger}\mathbf{H}^{\dagger}\mathbf{H}\mathbf{f}. \tag{5}$$

When perfect CSI ($\mathbf{H} = \mathsf{H}$) is also available at the transmitter, the optimal choice ($\mathbf{f}_{\mathsf{opt}}$) of beamforming vector on $\mathcal{G}(N_t, 1)$ that maximizes the received $\mathsf{SNR}$ is $\mathbf{u}_{\mathsf{H}}$, the dominant right singular vector of $\mathsf{H}$ (which is also the dominant eigenvector of $\mathsf{H}^{\dagger}\mathsf{H}$). In this case, the received $\mathsf{SNR}$ is given by $\rho\lambda_1$, where $\lambda_1$ is the dominant eigenvalue of $\mathsf{H}^{\dagger}\mathsf{H}$.

However, perfect CSI is hard to obtain at the transmitter end in practice. Thus, as motivated in Sec. I, we assume a $B$-bit limited-feedback model for the reverse link. We need the following definition to introduce the codebook model.

*Definition 1* ***(Exchangeable & Isotropic random variables):*** A family of random variables, $X_1, \cdots, X_n$, is said to be *exchangeable* if the joint distribution is invariant to the set of permutations over $\{1, \cdots, n\}$. That is,

$$\Pr\big(X_1, \cdots, X_n \in \Theta\big) = \Pr\big(X_{\pi_1}, \cdots, X_{\pi_n} \in \Theta\big) \tag{6}$$

for all permutations $\Pi = [\pi_1, \cdots, \pi_n]$ and any $\Theta$ in the range space of $\{X_1, \cdots, X_n\}$. A family of i.i.d. random variables is exchangeable. Exchangeable random variables are identically distributed [40].

A random $N_t \times 1$ unit-norm vector $\mathbf{f}$ is said to be *isotropic* if its distribution is invariant to pre- and post-multiplication by unitary matrices. That is,

$$\Pr\big(\mathbf{f} \in \Theta\big) = \Pr\big(e^{j\phi}\mathbf{U}\,\mathbf{f} \in \Theta\big) \tag{7}$$

for all $N_t \times N_t$ unitary matrices $\mathbf{U}$ and $\phi \in [0, 2\pi)$, and $\Theta$ in the range space $\mathcal{G}(N_t, 1)$. In particular, the distribution function of an $N_t \times 1$ isotropic beamforming vector is given as [41]

$$\Pr\big(\mathbf{f} \in \Theta\big) = \int_{\theta \in \Theta} \frac{\Gamma(N_t)}{\pi^{N_t}} \cdot \delta\big(\mathbf{f}^{\dagger}\mathbf{f} - 1\big)\, d\theta \tag{8}$$

where $\delta(\cdot)$ stands for the Dirac delta operator and

$$\Gamma(x) = \int_0^{\infty} t^{x-1} e^{-t} dt \tag{9}$$

stands for the Gamma function extended to $\mathbb{C}$ (minus its singularities). ■

In this work, we assume that an RVQ codebook of $B$ bits, $\mathcal{C} = \{\mathbf{f}_i, i = 1, \cdots, 2^B\}$, is known *a priori* at both the ends. The beamforming vectors in $\mathcal{C}$ are isotropic and i.i.d. over $\mathcal{G}(N_t, 1)$. The index $i^{\star}$ of the codeword that maximizes the received SNR,

$$i^{\star} = \arg\max_i \ \mathbf{f}_i^{\dagger}\mathbf{H}^{\dagger}\mathbf{H}\mathbf{f}_i, \tag{10}$$

is fed back using $B$ bits. We assume that there is no error or delay in feeding the index back.

Since an RVQ codebook is by construction random, our interest is in the average properties of an ensemble of RVQ codebooks. We desire to compute the following quantities:

$$\Delta\mathsf{SNR}_{\mathsf{rx}} \triangleq \mathbf{E}_{\mathcal{C}}\left[\mathbf{E}_{\mathbf{H}}\left[\frac{\lambda_1 - \max_i \mathbf{f}_i^{\dagger}\mathbf{H}^{\dagger}\mathbf{H}\mathbf{f}_i}{\lambda_1}\right]\right] \tag{11}$$

$$\Delta I \triangleq \mathbf{E}_{\mathcal{C}}\Big[\mathbf{E}_{\mathbf{H}}\big[I_{\mathsf{perf}} - I_{\mathsf{lim}}\big]\Big]. \tag{12}$$

The received SNR loss, $\Delta\mathsf{SNR}_{\mathsf{rx}}$, is the ensemble average (over the family of RVQ codebooks) of the average (over channel randomness) normalized received SNR loss relative to a perfect CSI scheme. The quantity $\Delta I$ is the ensemble average of the loss in average mutual information. In (12), $I_{\mathsf{perf}}$ and $I_{\mathsf{lim}}$ denote the mutual information[2] achievable with channel realization $\mathbf{H} = \mathsf{H}$ with perfect CSI and limited-feedback using the feedback metric in (10), respectively:

$$I_{\mathsf{perf}} = \log\left(1 + \rho \cdot \lambda_1\right) \tag{13}$$

$$I_{\mathsf{lim}} = \log\Big(1 + \rho \cdot \max_i \mathbf{f}_i^{\dagger}\mathsf{H}^{\dagger}\mathsf{H}\mathbf{f}_i\Big) \tag{14}$$

where $\lambda_1 \geq \cdots \geq \lambda_{N_t}$ are the eigenvalues of $\mathsf{H}^{\dagger}\mathsf{H}$ in decreasing order.

[2]All logarithms are to base 2, unless specified otherwise.

## III. Received SNR Loss

The goal of this section is to produce a tractable characterization of $\Delta\mathsf{SNR}_{\mathsf{rx}}$ as defined in (11). For this, note that a simple Fubini argument implies that we can change the order of expectation in (11) (and (12)). Thus, conditioned on a particular realization of the channel $\mathbf{H} = \mathsf{H}$, we seek to compute the following average:

$$\mathbf{E}_{\mathcal{C}} \left[ \frac{\lambda_1 - \max_i \mathbf{f}_i^\dagger \mathsf{H}^\dagger \mathsf{H} \mathbf{f}_i}{\lambda_1} \right] \triangleq \Delta_1. \tag{15}$$

We then average $\Delta_1$ over $\mathbf{H}$ to obtain $\Delta\mathsf{SNR}_{\mathsf{rx}}$.

### A. Equivalent Characterization of $\Delta_1$

*Lemma 1:*

- If $\{\mathbf{f}_i\}$ are isotropic on $\mathcal{G}(N_t, 1)$, the family of random variables
$$\left\{ |\mathbf{f}_i(k)|^2,\, k = 1, \cdots, N_t \right\} \tag{16}$$
is exchangeable for any fixed $i$. Recall that $\mathbf{f}_i(k)$ is the $k$-th element of $\mathbf{f}_i$.
- Further, with a given fixed channel realization $\mathbf{H} = \mathsf{H}$, the family of random variables $\{\mathbf{x}_i,\, i = 1, \cdots, 2^B\}$ where $\mathbf{x}_i = \mathbf{f}_i^\dagger \mathsf{H}^\dagger \mathsf{H} \mathbf{f}_i$ is i.i.d. over its range $[\lambda_{N_t}, \lambda_1]$.

*Proof:* See Appendix A. ■

If $\mathbf{x}_i$ are i.i.d. positive random variables, for any $x > 0$, we have

$$\Pr\left( \max_{i=1,\cdots,m} \mathbf{x}_i \leq x \right) = \Big( \Pr\left( \mathbf{x}_i \leq x \right) \Big)^m \tag{17}$$

for any choice of $m$. Using this fact in conjunction with Lemma 1, we have

$$\mathbf{E}_{\mathcal{C}} \left[ \max_i \mathbf{f}_i^\dagger \mathsf{H}^\dagger \mathsf{H} \mathbf{f}_i \right] - \lambda_{N_t} = \int_{\lambda_{N_t}}^{\lambda_1} \Pr\left( \max_i \mathbf{f}_i^\dagger \mathsf{H}^\dagger \mathsf{H} \mathbf{f}_i > x \right) dx \tag{18}$$

$$= \lambda_1 - \lambda_{N_t} - \int_{\lambda_{N_t}}^{\lambda_1} \Pr\left( \max_i \mathbf{f}_i^\dagger \mathsf{H}^\dagger \mathsf{H} \mathbf{f}_i \leq x \right) dx \tag{19}$$

where (18) follows from a routine Fubini argument. Hence, upon rearrangement, we have

$$\Delta_1 = \frac{1}{\lambda_1} \cdot \left( \lambda_1 - \mathbf{E}_{\mathcal{C}} \left[ \max_i \mathbf{f}_i^\dagger \mathsf{H}^\dagger \mathsf{H} \mathbf{f}_i \right] \right) \tag{20}$$

$$= \frac{1}{\lambda_1} \cdot \int_{\lambda_{N_t}}^{\lambda_1} \Big( \Pr\left( \mathbf{f}^\dagger \mathsf{H}^\dagger \mathsf{H} \mathbf{f} \leq x \right) \Big)^m dx \tag{21}$$

$$= \frac{1}{\lambda_1} \cdot \int_{\lambda_{N_t}}^{\lambda_1} \Big( \Pr\left( \mathbf{f}^\dagger \mathsf{\Lambda} \mathbf{f} \leq x \right) \Big)^m dx \tag{22}$$

where the eigen-decomposition of $\mathsf{H}^\dagger \mathsf{H}$ is given as $\mathsf{H}^\dagger \mathsf{H} = \mathsf{U} \mathsf{\Lambda} \mathsf{U}^\dagger$ with $\mathsf{\Lambda} = \mathsf{diag}\left( \left[ \lambda_1, \cdots, \lambda_{N_t} \right] \right)$, $\mathbf{f}$ is an isotropically distributed vector in $\mathcal{G}(N_t, 1)$ in (21) and (22), and $m$ is particularized to $m = 2^B$ in (21) and (22).

### B. Distribution Function of the Weighted-Norm of Unit-Norm Vectors

From the preceding discussion, we conclude that computation of $\Delta\mathsf{SNR}_{\mathsf{rx}}$ requires the distribution function of $\mathbf{f}^{\dagger}\Lambda\mathbf{f}$, which is a weighted-norm (with weights given by the diagonal entries of $\Lambda$) of isotropically distributed beamforming vectors on $\mathcal{G}(N_t, 1)$. We start by characterizing the relevant distribution functions completely in the special cases of $N_t = 2, 3$. (A study of the general $N_t$ case follows.)

*Lemma 2:* Let $\mathbf{f}$ be an isotropically distributed unit-norm vector on $\mathcal{G}(N_t, 1)$ and let $\Lambda = \mathsf{diag}\left([\lambda_1, \cdots, \lambda_{N_t}]\right)$ be some fixed diagonal matrix with $\lambda_1 \geq \cdots \geq \lambda_{N_t} \geq 0$. The cumulative distribution function (CDF) $F(x)$ of $\mathbf{f}^{\dagger}\Lambda\mathbf{f}$ over the non-trivial support region (the interval $[\lambda_{N_t}, \lambda_1]$) is as follows:

$$F(x)\Big|_{N_t=2} = \frac{x-\lambda_2}{\lambda_1-\lambda_2}, \quad \lambda_2 \leq x \leq \lambda_1, \tag{23}$$

$$F(x)\Big|_{N_t=3} = \begin{cases} \frac{(x-\lambda_3)^2}{(\lambda_1-\lambda_3)(\lambda_2-\lambda_3)}, & \lambda_3 \leq x \leq \lambda_2 \\ F(\lambda_2) + \frac{(x-\lambda_2)(2\lambda_1-x-\lambda_2)}{(\lambda_1-\lambda_2)(\lambda_1-\lambda_3)}, & \lambda_2 < x \leq \lambda_1. \end{cases} \tag{24}$$

While the behavior of $F(x)$ is too cumbersome to be stated in the general $N_t$ case, its behavior over the segment $[\lambda_2, \lambda_1]$ is simple:

$$F(x) = 1 - \frac{(\lambda_1 - x)^{N_t-1}}{\prod_{j=2}^{N_t}(\lambda_1 - \lambda_j)}, \ \lambda_2 \leq x \leq \lambda_1. \tag{25}$$

*Proof:* See Appendix B. ■

A simple verification shows that $F(\lambda_1) = 1$ in all the cases, as expected. The distribution functions are derived in Appendix B by computing the volume of intersection of a complex ellipsoid with a unit-radius complex sphere. This computation mirrors and generalizes the computation in [8] where the volume of a spherical cap (intersection of a plane with a unit-radius complex sphere) is obtained in closed-form. While this generalization is hard to geometrically visualize beyond the $N_t = 2$ case, it can be seen that the trend over $[\lambda_2, \lambda_1]$ shows the same behavior as the distribution function in [8].

Fig. 1 illustrates the trends of the CDF by plotting the goodness-of-fit between the theoretical expressions in Lemma 2 and the CDF estimated via Monte Carlo methods. Three cases are considered: a) $\Lambda = \mathsf{diag}([2 \ 1])$ for $N_t = 2$, b) $\Lambda = \mathsf{diag}([3 \ 2 \ 1])$ for $N_t = 3$, and c) $\Lambda = \mathsf{diag}([4 \ 3 \ 2 \ 1])$ for $N_t = 4$.

### C. Main Result

The following theorem captures the performance loss with RVQ codebooks.

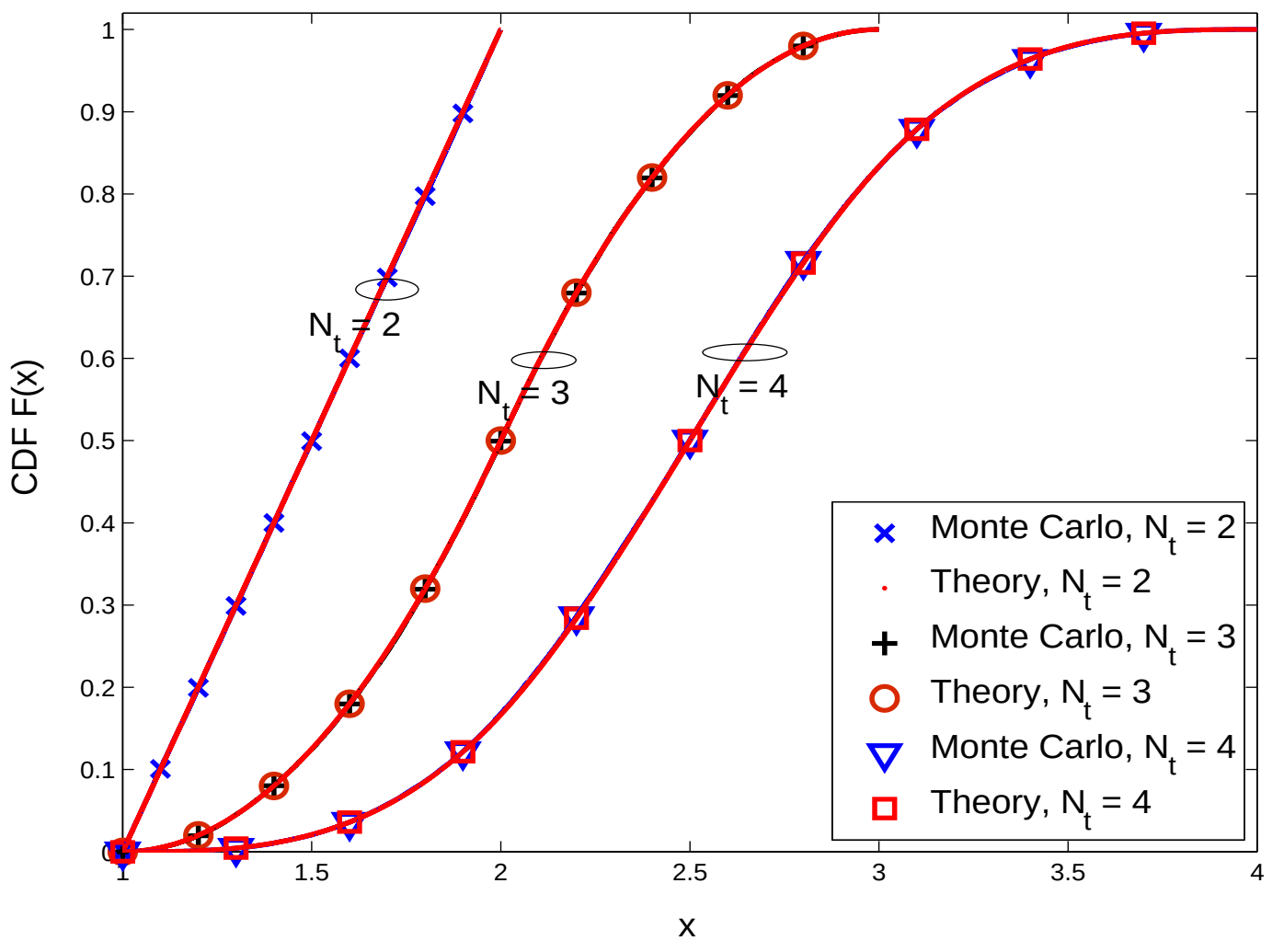

Fig. 1. CDF of weighted-norm of isotropically distributed unit-norm vectors.

*Theorem 1:* In the MIMO setting, in the special cases of $N_t = 2$ and $3$, we have

$$\Delta_1\Big|_{N_t=2} = A_2 \cdot \left[1 - \frac{\lambda_2}{\lambda_1}\right] \tag{26}$$

$$\Delta_1\Big|_{N_t=3} = A_3 \cdot \left[\left(1 - \frac{\lambda_3}{\lambda_1}\right)\left(\frac{\lambda_2 - \lambda_3}{\lambda_1 - \lambda_3}\right)^m + \left(1 - \frac{\lambda_2}{\lambda_1}\right) \times \sum_{k=1}^{m}\left(\frac{\lambda_2 - \lambda_3}{\lambda_1 - \lambda_3}\right)^{m-k} \frac{2^k m(m-1)\cdots(m-k+1)}{(2m-1)(2m-3)\cdots(2m-2k+1)}\right], \tag{27}$$

where $m = 2^B$, $A_2 = \frac{1}{2^B+1}$ and $A_3 = \frac{1}{2^{B+1}+1}$. In the general ($N_t \geq 4$) case, we have

$$\Delta_1 \approx A_{N_t} \cdot \left[\left(1 - \frac{\lambda_2}{\lambda_1}\right) \times \left[D^m + \sum_{k=1}^{m} \frac{2^k \cdot m(m-1)\cdots(m-k+1)}{(2m+p-1)\cdots(2m+p-2k+1)} D^{m-k}\right]\right] \triangleq \Delta_{1,\,\mathsf{appx}}, \tag{28}$$

$$A_{N_t} = \frac{1}{m(N_t-1)+1}, \quad p = \frac{2}{N_t - 1} - 1, \quad \text{and} \quad D \triangleq 1 - \prod_{j=2}^{N_t} \frac{\lambda_1 - \lambda_2}{\lambda_1 - \lambda_j}. \tag{29}$$

Further, we have the following bounds:

$$0 \leq \frac{\Delta_1 - \Delta_{1,\,\mathsf{appx}}}{\Delta_1} \leq \epsilon_B \tag{30}$$

where

$$\epsilon_B \triangleq \frac{\lambda_2 - \lambda_{N_t}}{\lambda_1} \cdot \frac{D^m}{\Delta_{1,\,\mathsf{appx}}}. \tag{31}$$

We will show subsequently (see (38)-(40)) that $\epsilon_B \stackrel{B\to\infty}{\to} 0$ for any $\mathsf{H}$. That is, $\Delta_{1,\,\mathsf{appx}}$ is a tight approximation to $\Delta_1$ with

$$\Delta_1 = \Delta_{1,\,\mathsf{appx}} + o\left(\Delta_{1,\,\mathsf{appx}}\right) \tag{32}$$

as $B \to \infty$.

*Proof:* Since $F(x)$ is monotonic, the dominant term of the integral in (22) in the general $N_t$ case is over the interval $[\lambda_2,\, \lambda_1]$. Computation of this dominant term results in the statement of the theorem. See Appendix C for details. ■

In the special cases where $\mathsf{H}$ is a MISO channel ($N_r = 1$) or $\mathsf{H}$ is effectively a MISO channel ($\mathsf{rank}(\mathsf{H}^\dagger\mathsf{H}) = 1$), $\Delta_1$ can be computed in closed-form [30, Cor. 1], [15] as

$$\Delta_1 = \mathbf{E}_{\mathcal{C}}\left[\min_i \sin^2(\theta_i)\right] = 2^B \beta\left(2^B, \frac{N_t}{N_t - 1}\right) \tag{33}$$

with $\theta_i$ denoting the angle between $\mathbf{f}_i$ and $\mathbf{u}_{\mathsf{H}}$ (the dominant right singular vector of $\mathsf{H}$) and

$$\beta(x, y) = \int_0^1 t^{x-1}(1-t)^{y-1} dt \tag{34}$$

is the Beta function. The MISO setting can be obtained as a limiting case of Theorem 1 with $\lambda_2 = \cdots = \lambda_{N_t} \to 0$.

## D. Asymptotics of $B$

Theorem 1 separates (to first order) the impact of the channel from that of the RVQ codebook (number of bits $B$). Nevertheless, the expressions provided are too complicated to obtain simple heuristic insights.

To overcome this difficulty, we now provide simplifications for $\Delta_1$ as $B \to \infty$. In the $N_t = 2$ setting, the expression for $\Delta_1$ is already simple. Thus, we start with the case of $N_t = 3$ and then study the $N_t \geq 4$ case.

*Proposition 1:* In the $N_t = 3$ case, the dominant term of $\Delta_1$ behaves as

$$\Delta_1 = \frac{\sqrt{\pi}}{2^{B/2+1}} \cdot \left[\left(1 - \frac{\lambda_2}{\lambda_1}\right)\left(1 + \frac{\lambda_2 - \lambda_3}{2(\lambda_1 - \lambda_3)}\right)\right] + o\left(2^{-B/2}\right) \tag{35}$$

as $B \to \infty$. Similarly, in the $N_t \geq 4$ case, we have

$$\Delta_1 = \underbrace{\frac{\kappa \cdot 2^{-\frac{B}{N_t-1}}}{N_t - 1}\left[\left(1 - \frac{\lambda_2}{\lambda_1}\right)\left(1 + \frac{D}{(1-D)(N_t-1)}\right)\right]}_{\Delta_{1,\,\mathsf{asymp}}} + o\left(2^{-\frac{B}{N_t-1}}\right), \tag{36}$$

where $\kappa = \Gamma\left(\frac{1}{N_t-1}\right)$ and $D$ is as in (29).

*Proof:* See Appendix D. ■

From Prop. 1 as well as (33), in the special case where $\mathsf{rank}(\mathsf{H}^\dagger \mathsf{H}) = 1$, we have

$$\Delta_1 = \frac{2^{-\frac{B}{N_t-1}}}{N_t - 1} + o\left(2^{-\frac{B}{N_t-1}}\right), \tag{37}$$

which is also established in [15], [30]. For the rate of convergence of $\epsilon_B$ in (31) as $B \to \infty$, note that

$$\log(\epsilon_B) = \log\left(\frac{\lambda_2 - \lambda_{N_t}}{\lambda_1}\right) + 2^B \log(D) + \log\left(\frac{1}{\Delta_{1,\,\mathsf{appx}}}\right) \tag{38}$$

$$\stackrel{(a)}{=} \frac{B}{N_t - 1} - 2^B \log\left(\frac{1}{D}\right) + \mathcal{O}(1) \tag{39}$$

$$\stackrel{B\to\infty}{\asymp} -2^B \log\left(\frac{1}{D}\right) \tag{40}$$

where (a) follows from Prop. 1 and the $\mathcal{O}(1)$ factor is a constant for a given $\mathsf{H}$.

We now provide a numerical study to illustrate the theoretical results presented in Theorem 1, and to provide an idea as to how useful the asymptotic approximations are in the non-asymptotic regime. Three channel realizations of size $N_r \times N_t$ with $N_t = N_r = \{2, 3, 4\}$ are generated randomly and then held constant and the performance is averaged over $1000$ RVQ codebooks. The three channels are such that the squared singular values are: 1) $[2\ \ 1]$, 2) $[3\ \ 2\ \ 1]$, and 3) $[4\ \ 3\ \ 2\ \ 1]$, respectively. Fig. 2 shows the match between the theoretical expressions in Theorem 1, the asymptotic approximations in Prop. 1 and Monte Carlo estimates of $\Delta_1$. We see that the asymptotic approximations are close even for small values of $B$ ($B \geq 2$), which is useful from a practically motivated limited-feedback perspective. While we have considered the goodness-of-fit of the three expressions with a specific channel realization in Fig. 2, the goodness-of-fit of the three expressions across a large family of channels is studied next.

## IV. Ordering Channels Based on RVQ Performance

The focus of this section is to develop a basis (or a metric) to *order* a family of channels such that the RVQ performance over a particular channel can be compared with performance over another channel. In particular, the interest is on those conditions on channels $\mathsf{H}_1$ and $\mathsf{H}_2$ that are critical to ensure that

$$\Delta_1\Big|_{\mathsf{H}_1} \leq \Delta_1\Big|_{\mathsf{H}_2}. \tag{41}$$

Let $\underline{\boldsymbol{\lambda}} = [\lambda_1, \cdots, \lambda_{N_t}]$ and $\underline{\boldsymbol{\mu}} = [\mu_1, \cdots, \mu_{N_t}]$ denote the vectors of squared singular values of $\mathsf{H}_1$ and $\mathsf{H}_2$ with $\lambda_1 \geq \cdots \geq \lambda_{N_t} \geq 0$ and $\mu_1 \geq \cdots \geq \mu_{N_t} \geq 0$. In the special case of $N_t = 2$, Theorem 1 shows that

$$\Delta_1\Big|_{\mathsf{H}_1} \leq \Delta_1\Big|_{\mathsf{H}_2} \iff \frac{\lambda_1}{\lambda_2} \leq \frac{\mu_1}{\mu_2}. \tag{42}$$

With $\underline{\boldsymbol{\lambda}}$ and $\underline{\boldsymbol{\mu}}$ normalized such that

$$\lambda_1 + \lambda_2 = \rho_c = \mu_1 + \mu_2, \tag{43}$$

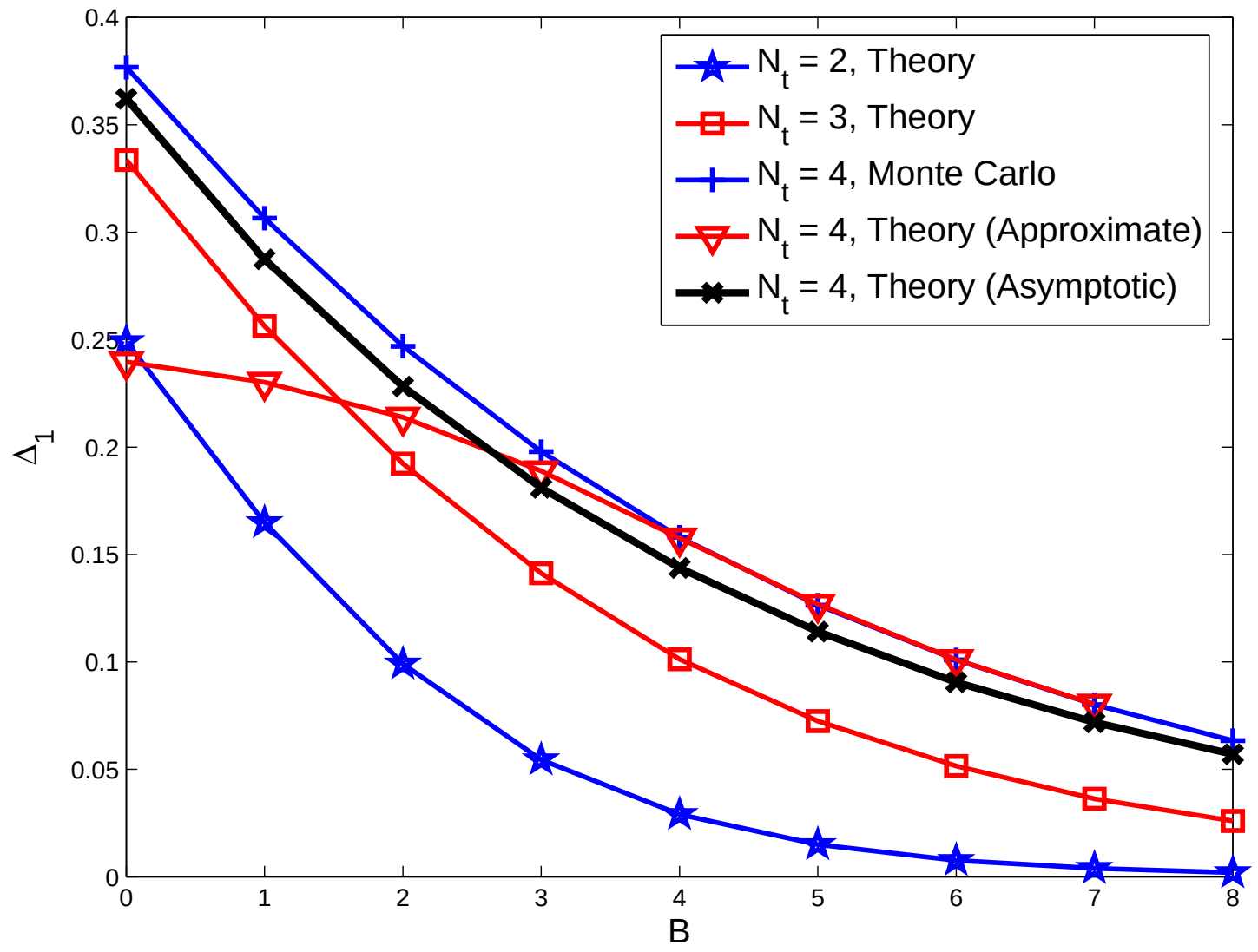

Fig. 2. Goodness-of-fit of different estimates of $\Delta_1$ as a function of $B$.

(42) is equivalent to $\lambda_1 \leq \mu_1$ or $\lambda_2 \geq \mu_2$. To make this connection more precise in the general $N_t$ case, we assume that the channels are normalized such that

$$\sum_{i=1}^{N_t} \lambda_i = \mathsf{Tr}(\mathsf{H}_1^\dagger \mathsf{H}_1) = \mathsf{Tr}(\mathsf{H}_2^\dagger \mathsf{H}_2) = \sum_{i=1}^{N_t} \mu_i = \rho_c, \tag{44}$$

where $\rho_c$ denotes the channel power. This normalization is commonly used in multi-antenna channel measurement studies to ensure that the channel power stays fixed, independent of the distance between the transmitter and the receiver and the energy of the scattering phenomena. See [39] for a discussion of channel power normalization issues.

We also define the notions of a majorization ordering and a Schur-convex function [42].

*Definition 2 (**Schur-convex function**):* We say that $\underline{\boldsymbol{\lambda}}$ is *majorized by* $\underline{\boldsymbol{\mu}}$ (denoted as $\underline{\boldsymbol{\lambda}} \prec \underline{\boldsymbol{\mu}}$) if

$$\sum_{i=1}^{k} \lambda_i \leq \sum_{i=1}^{k} \mu_i, \ 1 \leq k \leq N_t, \tag{45}$$

with equality for $k = N_t$. With $\underline{\boldsymbol{\lambda}}$ and $\underline{\boldsymbol{\mu}}$ denoting the vectors of squared singular values of $\mathsf{H}_1$ and $\mathsf{H}_2$, respectively, equality in (45) for $k = N_t$ is a consequence of (44).

Let $f(\cdot)$ be a function such that $f : \mathbb{R}^+_{N_t} \mapsto \mathbb{R}$. We say that $f(\cdot)$ is Schur-convex on $\mathbb{R}^+_{N_t}$ if

$$\mathbf{x} \prec \mathbf{y} \Longrightarrow f(\mathbf{x}) \leq f(\mathbf{y}). \tag{46}$$

The function $f(\cdot)$ is Schur-concave if $-f(\cdot)$ is Schur-convex. ■

With this background, the main result of this section is as follows.

*Theorem 2:* The normalized received $\mathsf{SNR}$ loss is a Schur-convex function of the squared singular values of the channel. That is, if $\underline{\boldsymbol{\lambda}}$ and $\underline{\boldsymbol{\mu}}$ denote the vectors of squared singular values of $\mathsf{H}_1$ and $\mathsf{H}_2$ with $\underline{\boldsymbol{\lambda}} \prec \underline{\boldsymbol{\mu}}$, we have

$$\Delta_1 \Big|_{\mathsf{H}_1} \leq \Delta_1 \Big|_{\mathsf{H}_2}. \tag{47}$$

*Proof:* See Appendix E. ■

Some comments are in order at this stage.

1) Note that it is difficult to draw the conclusion of Theorem 2 from either the exact expression in the $N_t = 3$ case or the approximate/asymptotic expressions of Sec. III. Theorem 2 provides a continuous ordering on the space of all possible (majorizable) channels with respect to RVQ performance. Similar results exploiting majorization theory have been obtained for the ergodic capacity of MISO systems [43], outage probability of MISO systems, error performance of orthogonal space-time block codes, performance analysis of precoding in MIMO systems [44], performance of CDMA systems, etc., (see [21], [44], [45] for details). Theorem 2 leads us to the following conclusion.

   *Corollary 1:* Any channel $\mathsf{H}$ with the vector of squared singular values denoted by $\underline{\boldsymbol{\lambda}}$ satisfies

$$\left[ \frac{\rho_c}{N_t}, \cdots, \frac{\rho_c}{N_t} \right] \prec \underline{\boldsymbol{\lambda}} \prec \left[ \rho_c, 0, \cdots, 0 \right] \tag{48}$$

   resulting in

$$\Delta_1 \Big|_{\left[ \frac{\rho_c}{N_t}, \cdots, \frac{\rho_c}{N_t} \right]} \leq \Delta_1 \Big|_{\underline{\boldsymbol{\lambda}}} \leq \Delta_1 \Big|_{\left[ \rho_c, 0, \cdots, 0 \right]}. \tag{49}$$

   In other words, the best channel with respect to RVQ performance is well-conditioned with squared condition number $\chi_{\mathsf{H}} = \frac{\lambda_1(\mathsf{H}^\dagger \mathsf{H})}{\lambda_{N_t}(\mathsf{H}^\dagger \mathsf{H})}$ equal to $1$, whereas the worst channel is a $\mathsf{rank}$-$1$ channel. ■

   This conclusion fits within the theme of *source-channel matching* for signalling design in single-user MIMO systems, established in [21]: the best channel with respect to a specific signalling scheme is the channel that optimizes an appropriately defined *matching metric* for that scheme. For the beamforming scheme with $\Delta_1$ as the chosen metric and given that an RVQ codebook has isotropic vectors (equally likely to beamform along any direction), the channel that is best-suited to this scheme should also have dominant right singular vectors that are isotropic in $\mathcal{G}(N_t, 1)$. This choice leads us to the i.i.d. channel matrix [7], [10]. Similarly, a $\mathsf{rank}$-$1$ channel with a fixed right singular vector is ill-suited to an RVQ codebook that is "wasteful" by beamforming isotropically in $\mathcal{G}(N_t, 1)$.

2) We now provide two specific examples to illustrate the dependence of $\Delta_1$ on the rank of the channel and the condition number.

*Corollary 2:* Note that

$$[\rho_c/N_t, \cdots, \rho_c/N_t] \prec \cdots \prec \left[ \underbrace{\rho_c/r}_{r \text{ times}}, \underbrace{0}_{N_t - r \text{ times}} \right] \prec \cdots \prec [\rho_c, 0, \cdots, 0]. \tag{50}$$

Thus, $\Delta_1$ increases as the rank $r$ of the channel decreases.

Further, within the family of channels with the same rank $r$, $\Delta_1$ increases as the $r$ non-zero squared singular values become more ill-conditioned. ■

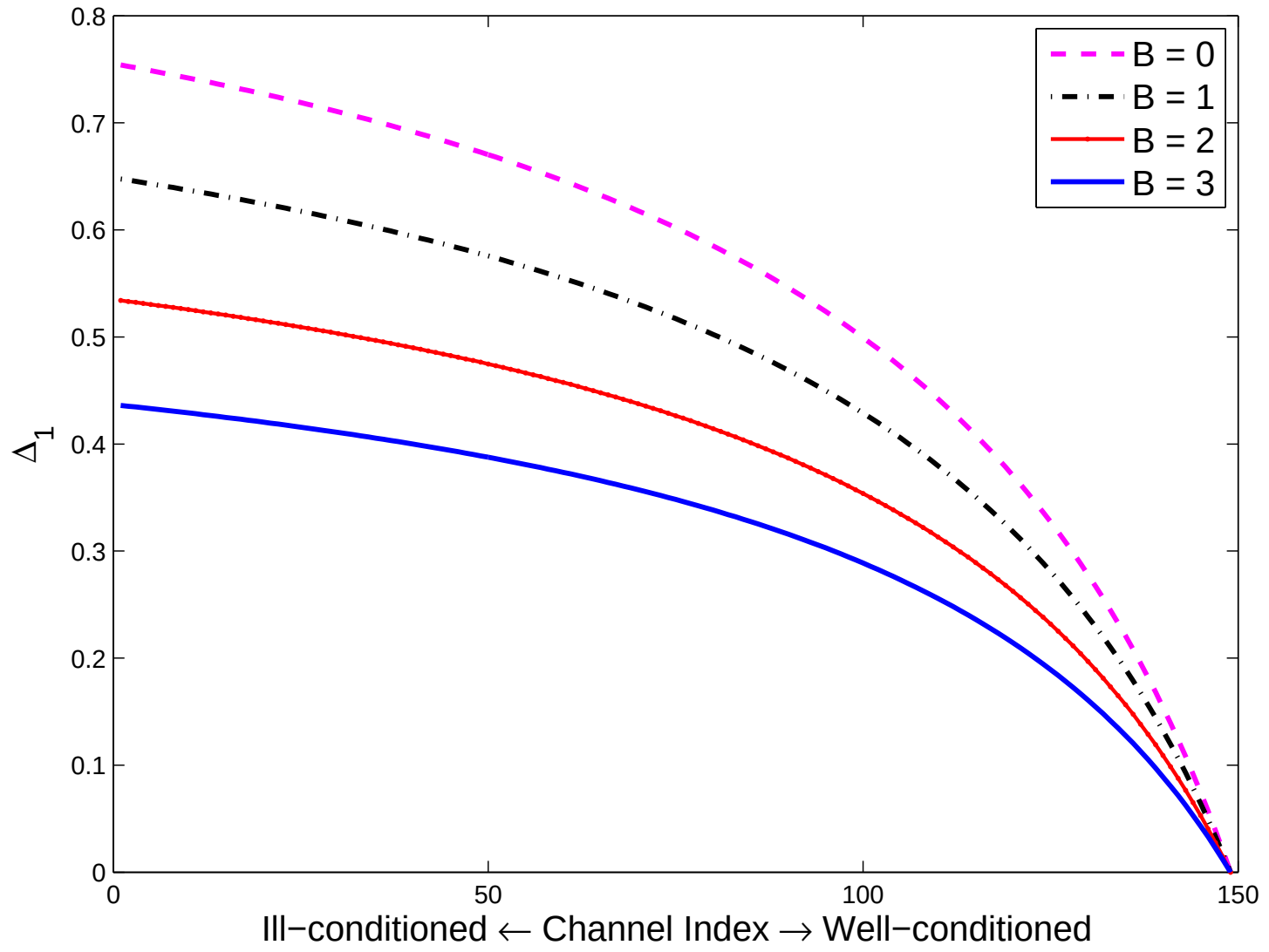

Fig. 3. Received SNR loss for channels ordered via a majorization relationship as a function of $B$.

Fig. 3 plots $\Delta_1$ as a function of $B$ across a family of $150$ channels that can be continuously majorized as follows. With $N_t = N_r = 4$ and $\rho_c$ set arbitrarily to $1$ (without loss in generality), the squared singular values for the $i$-th channel are given as

$$\underline{\boldsymbol{\lambda}}_i \triangleq [1 - x_i, \, x_i/3, \, x_i/3, \, x_i/3] \tag{51}$$

where $x_i$ increases from $0.01$ to $0.75$ in steps of $0.005$. It can be seen that for any $i$

$$\underline{\boldsymbol{\lambda}}_i \prec \underline{\boldsymbol{\lambda}}_j, \, 1 \leq j \leq i - 1, \tag{52}$$

and the channel becomes more well-conditioned as $i$ increases. On the other hand, $\Delta_1$ continuously decreases, thus illustrating Theorem 2.

3) Majorization provides an ordering metric to compare channels with respect to RVQ performance. However, it is important to note that the metric only induces a *partial* ordering on the family of channels since there exist channels that cannot be compared via a majorization relationship. A simplified, albeit approximate, channel ordering metric that reflects the condition number of the channel and allows an *approximate complete* ordering of channels is $\lambda_1$. However, numerical results illustrating the efficacy of this metric are not provided here for the sake of brevity.

In general, we would like to study the behavior of $\Delta \mathsf{SNR}_{\mathsf{rx}} = \mathbf{E}_{\mathbf{H}} \left[ \Delta_1 \right]$.

*Proposition 2:* In the special case where $\{N_t, N_r\} \rightarrow \infty$ with $\frac{N_t}{N_r} \rightarrow 0$, the singular values of $\mathbf{H}$ converge (harden) [21], [46] as follows:

$$\lambda_i(\mathbf{H}^\dagger \mathbf{H}) \rightarrow \lambda_i \left( \mathbf{E} \left[ \mathbf{H}^\dagger \mathbf{H} \right] \right) = \lambda_i \left( \mathbf{\Sigma}_t \right), \ i = 1, \cdots, N_t. \tag{53}$$

Hence, we have

$$\Delta \mathsf{SNR}_{\mathsf{rx}} \overset{\mathcal{O}(1)}{\approx} \underbrace{\frac{\lambda_1(\mathbf{\Sigma}_t) - \lambda_2(\mathbf{\Sigma}_t)}{\lambda_1(\mathbf{\Sigma}_t)}}_{\mathcal{D}_1} \cdot \underbrace{\left( 1 + \prod_{j=3}^{N_t} \frac{\lambda_1(\mathbf{\Sigma}_t) - \lambda_j(\mathbf{\Sigma}_t)}{\lambda_1(\mathbf{\Sigma}_t) - \lambda_2(\mathbf{\Sigma}_t)} \right)}_{\mathcal{D}_2} \tag{54}$$

with the approximation holding up to a multiplicative constant that depends on the antenna dimensions and $B$. ■

Note that $\mathcal{D}_1$ is minimized when $\lambda_1(\mathbf{\Sigma}_t) \approx \lambda_2(\mathbf{\Sigma}_t)$ whereas $\mathcal{D}_2$ is minimized when $\lambda_2(\mathbf{\Sigma}_t) \approx \cdots \approx \lambda_{N_t}(\mathbf{\Sigma}_t) \approx 0$. But $\mathcal{D}_1 \mathcal{D}_2$ is minimized when $\mathbf{\Sigma}_t$ is well-conditioned. Apart from this case, estimating $\Delta \mathsf{SNR}_{\mathsf{rx}}$ appears to be difficult in general. We therefore resort to numerical studies to study trends of $\Delta \mathsf{SNR}_{\mathsf{rx}}$.

Following the discussion in the context of channel ordering, we expect that as the rank of $\mathbf{\Sigma}_t$ increases and as a consequence, the condition number of the channel decreases on average, the performance loss with RVQ should decrease. Fig. 4(a) illustrates this heuristic with four channels generated according to the Kronecker-product correlation model in (2). The eigenvalues of $\mathbf{\Sigma}_r$ of the four channels are fixed as $1.6 \times \begin{bmatrix} 4 & 3 & 2 & 1 \end{bmatrix}$ where the factor of $1.6$ means that $\mathsf{Tr}(\mathbf{\Sigma}_r) = N_t N_r = 16$. The eigenvalues of $\mathbf{\Sigma}_t$ are as follows: 1) $[16 \ 0 \ 0 \ 0]$, 2) $[8 \ 8 \ 0 \ 0]$, 3) $[16/3 \ 16/3 \ 16/3 \ 0]$, 4) $[4 \ 4 \ 4 \ 4]$ ensuring that $\mathsf{Tr}(\mathbf{\Sigma}_t) = 16$ in all the four cases.

## V. Mutual Information Loss

Following the same development as in Sec. III, we can write $\Delta I$ as

$$\Delta I = \mathbf{E}_{\mathbf{H}} \left[ \Delta_2 \right], \ \ \Delta_2 \ = \ \int_L^U \Big( \Pr \left( \mathbf{x} \leq x \right) \Big)^m dx \tag{55}$$

where $\mathbf{x} = \log \left( 1 + \rho \cdot \mathbf{f}^\dagger \mathsf{H}^\dagger \mathsf{H} \mathbf{f} \right)$, $m = 2^B$,

$$L = \log \left( 1 + \rho \lambda_{N_t} \right), \ \ \text{and} \ \ U = \log \left( 1 + \rho \lambda_1 \right). \tag{56}$$

It is easy to see that

$$\Delta_2 \ = \ \frac{\rho}{\log_e(2)} \cdot \int_{\lambda_{N_t}}^{\lambda_1} \frac{\left( \Pr \left( \mathbf{f}^\dagger \mathsf{H}^\dagger \mathsf{H} \mathbf{f} \leq x \right) \right)^m}{1 + \rho x} \, dx \tag{57}$$

$$= \ \frac{\rho}{\log_e(2)} \cdot \int_{\lambda_{N_t}}^{\lambda_1} \frac{\left( \Pr \left( \mathbf{f}^\dagger \mathsf{\Lambda} \mathbf{f} \leq x \right) \right)^m}{1 + \rho x} \, dx. \tag{58}$$

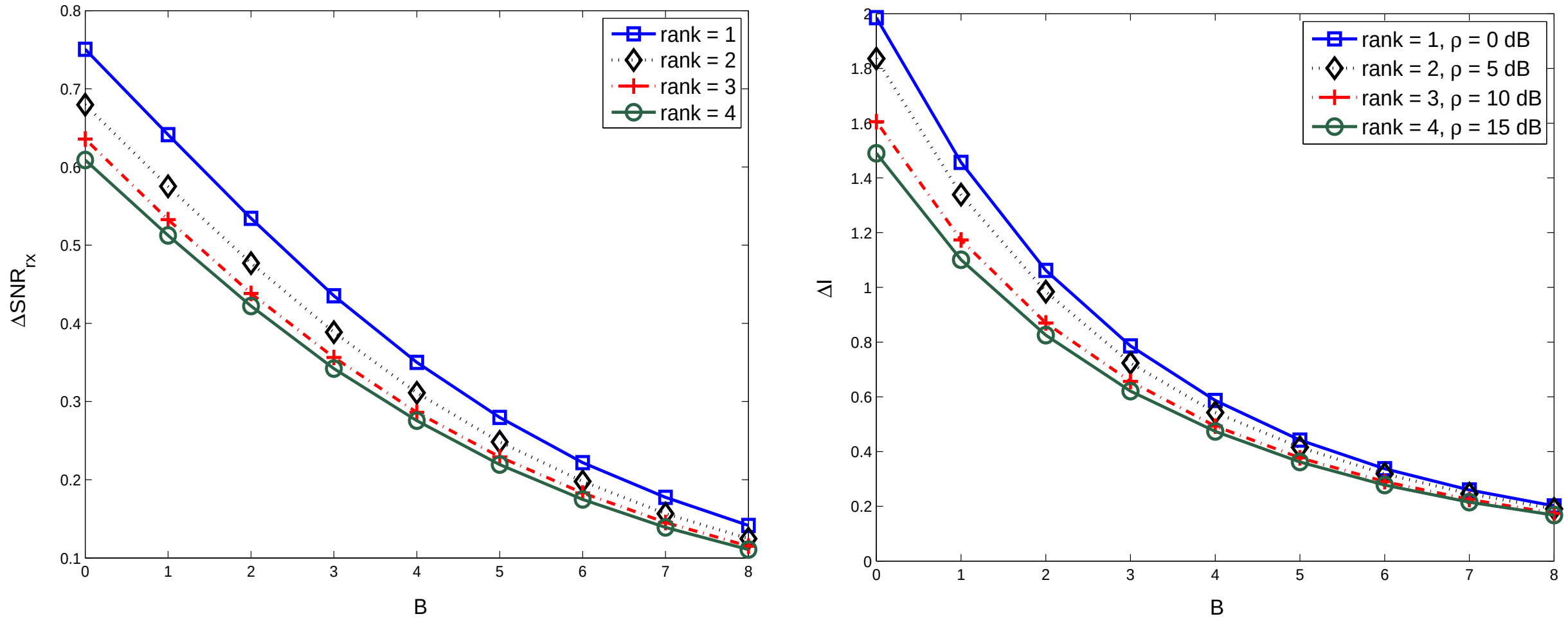

(a) (b)

Fig. 4. (a) Average received SNR loss and (b) Average mutual information loss as a function of the rank of $\mathbf{\Sigma}_t$.

In contrast to the development in Sec. III where the integrand is monotonically increasing, the integrand in (58) is not necessarily monotonic as it is a ratio of two increasing functions. Nevertheless, we can trivially capture the trend of $\Delta_2$ as illustrated next.

*Corollary 3:* The following asymptotic trend holds for $\Delta_2$:

$$\Delta_2 = 2^{-\frac{B}{N_t - 1}} \cdot \frac{\rho(\lambda_1 - \lambda_2) \cdot \kappa}{\log_e(2)(N_t - 1)} \cdot \left[ 1 + \frac{D}{(1-D)(N_t - 1)} \right] + o\left( 2^{-\frac{B}{N_t - 1}} \right) \tag{59}$$

where $\kappa$ and $D$ are as in Theorem 1.

*Proof:* A trivial bound for $1 + \rho x$ in (58) over the interval $[\lambda_2, \lambda_1]$ implies that the dominant term of $\Delta_2$ (denoted as $\Delta_{2,\,\mathsf{appx}}$) can be bounded as

$$\frac{\rho}{1 + \rho\lambda_1} \leq \frac{\Delta_{2,\,\mathsf{appx}} \cdot \log_e(2)}{\int_{\lambda_2}^{\lambda_1} \left( \Pr\left( \mathbf{f}^{\dagger} \mathbf{\Lambda} \mathbf{f} \leq x \right) \right)^m dx} \leq \frac{\rho}{1 + \rho\lambda_2}. \tag{60}$$

A consequence of the computation in Theorem 1 is that

$$\underline{\Delta}_2 \leq \Delta_{2,\,\mathsf{appx}} \leq \overline{\Delta}_2 \tag{61}$$

with

$$\underline{\Delta}_2 \triangleq \frac{\rho}{\log_e(2)} \cdot A_{N_t} \cdot \left( \frac{\lambda_1 - \lambda_2}{1 + \rho\,\lambda_1} \right) \cdot \mathsf{C}_1 \tag{62}$$

$$\overline{\Delta}_2 \triangleq \frac{\rho}{\log_e(2)} \cdot A_{N_t} \cdot \left( \frac{\lambda_1 - \lambda_2}{1 + \rho\,\lambda_2} \right) \cdot \mathsf{C}_1 \tag{63}$$

where

$$\mathsf{C}_1 = D^m + \sum_{k=1}^{m} \frac{2^k \cdot m(m-1)\cdots(m-k+1)}{(2m+p-1)\cdots(2m+p-2k+1)} D^{m-k} \tag{64}$$

and we have reused the notations $(A_{N_t}, p, D)$ from Theorem 1. It is straight-forward to see that

$$\frac{\underline{\Delta}_2}{\left(\frac{\rho(\lambda_1-\lambda_2)}{1+\rho\lambda_1}\right)} = \frac{\kappa \cdot 2^{-\frac{B}{N_t-1}}}{\log_e(2)(N_t-1)} \cdot \left[1 + \frac{D}{(1-D)(N_t-1)}\right] + o\left(2^{-\frac{B}{N_t-1}}\right) \tag{65}$$

$$\frac{\overline{\Delta}_2}{\left(\frac{\rho(\lambda_1-\lambda_2)}{1+\rho\lambda_2}\right)} = \frac{\kappa \cdot 2^{-\frac{B}{N_t-1}}}{\log_e(2)(N_t-1)} \cdot \left[1 + \frac{D}{(1-D)(N_t-1)}\right] + o\left(2^{-\frac{B}{N_t-1}}\right), \tag{66}$$

and thus we have (59). ■

While Cor. 3 captures the asymptotic trend of $\Delta_2$ via trivial bounding, it is not tight when $\lambda_1 \gg \lambda_2$. In these situations, it is useful to obtain a tighter estimate for $\Delta_2$. This is addressed next.

*Theorem 3:* In the $N_t = 2$ case, we have

$$\Delta_2 = \frac{1}{\log_e(2) \cdot z^m} \left[\log_e(1+z) - \sum_{t=1}^{m} \frac{(-1)^{t+1} z^t}{t}\right] \tag{67}$$

where $m = 2^B$ and $z \triangleq \frac{\rho(\lambda_1-\lambda_2)}{1+\rho\lambda_2}$. In the general $N_t$ case, we have the following approximations:

$$\Delta_2 \approx \frac{\frac{\rho A}{\log_e(2)}}{(N_t-1)(1+\rho\lambda_1)} \cdot \sum_{i=0}^{\infty} \frac{\gamma^i (1-D)^{\frac{i+1}{N_t-1}}}{m + \frac{i+1}{N_t-1}} \times \left[D^m + \sum_{k=1}^{m} \frac{2^k \cdot m(m-1)\cdots(m-k+1) \cdot D^{m-k}}{(2m+p_i-1)\cdots(2m+p_i-2k+1)}\right] \triangleq \Delta_{2,\,\mathsf{appx}}, \tag{68}$$

$$\gamma = \frac{\rho A}{1+\rho\lambda_1}, \quad A = \left(\prod_{j \geq 2} \lambda_1 - \lambda_j\right)^{\frac{1}{N_t-1}}, \quad \text{and} \quad p_i = \frac{2(i+1)}{N_t-1} - 1. \tag{69}$$

Further, we have

$$0 \leq \frac{\Delta_2 - \Delta_{2,\,\mathsf{appx}}}{\Delta_2} \leq \epsilon'_B \tag{70}$$

where

$$\epsilon'_B \triangleq \frac{\rho(\lambda_2 - \lambda_{N_t})}{(1+\rho\lambda_{N_t}) \cdot \log_e(2)} \cdot \frac{D^m}{\Delta_{2,\,\mathsf{appx}}} \tag{71}$$

$$\log(\epsilon'_B) \overset{B\to\infty}{\asymp} -2^B \log\left(\frac{1}{D}\right). \tag{72}$$

Thus $\epsilon'_B \overset{B\to\infty}{\to} 0$ and

$$\Delta_2 = \Delta_{2,\,\mathsf{appx}} + o\left(\Delta_{2,\,\mathsf{appx}}\right) \tag{73}$$

as $B \to \infty$.

*Proof:* See Appendix F. ■

An alternate expansion for $\Delta_2$ is also presented in Appendix F. This expansion corresponds to an alternate form of the integrand in (58) and is captured by a series where the signs of alternate terms change. In this spirit, the alternate expansion generalizes (67). From a numerical standpoint, this oscillatory nature is unattractive due to non-convergence of the series and (68)-(69) overcomes this problem. We now study the asymptotic trends of $\Delta_2$.

*Proposition 3:* In the $N_t = 2$ case, depending on the relationship between $\rho, \lambda_1$ and $\lambda_2$, two possibilities arise as $B$ increases. We have

$$\Delta_2 \overset{B \to \infty}{\asymp} \begin{cases} \frac{1}{\log_e(2)} \cdot \frac{z}{(m+1)}, & z < 1 \\ \frac{1}{\log_e(2)} \cdot \frac{(z-1)}{2 \cdot z \log_e(z) \cdot (m-1)}, & z \geq 1. \end{cases} \tag{74}$$

In the general $N_t$ case, as $B \to \infty$, we have

$$\Delta_2 = \underbrace{\frac{2^{-\frac{B}{N_t - 1}}}{\log_e(2)(N_t - 1)} \cdot \frac{\rho(\lambda_1 - \lambda_2)}{1 + \rho\lambda_1} \cdot \left[\kappa + \frac{D}{1 - D}\right]}_{\Delta_{2,\text{ asymp}}} + o\left(2^{-\frac{B}{N_t - 1}}\right) \tag{75}$$

where $\kappa = \Gamma\left(\frac{1}{N_t - 1}\right)$ and $D$ is as in (29).

*Proof:* See Appendix G. ■

We now illustrate the above theoretical results in Fig. 5(a) and (b) where we plot the instantaneous mutual information loss both theoretically and via Monte Carlo averaging. The squared singular values of the three channels are (as before): 1) $[2 \;\; 1]$ for $N_t = 2$, 2) $[3 \;\; 2 \;\; 1]$ for $N_t = 3$, and 3) $[4 \;\; 3 \;\; 2 \;\; 1]$ for $N_t = 4$, respectively. Asymptotic and approximate expressions are tight for small $B$ values as long as $\rho$ is not too large. On the other hand, Fig. 4(b) illustrates the trend of $\Delta I$ as a function of the rank of $\mathbf{\Sigma}_t$. The channel data used to generate Fig. 4(b) is the same as that used for generating Fig. 4(a) (see the discussion there).

## VI. Skewed Codebooks for Correlated Channels

From (22) and (55), the asymptotic optimality of RVQ codebooks in the correlated case is obvious. That is, $\Delta \mathsf{SNR}_{\mathsf{rx}} \to 0$ and $\Delta I \to 0$ (respectively) as $B \to \infty$, *independent* of the channel correlation profile, since a probability term in the integrand is raised to the power of $m = 2^B \to \infty$. Nevertheless, this does not mean that RVQ codebooks are optimal for any finite value of $B$ in the correlated case. While the ensemble averaging of RVQ codebooks is necessary to make constructive statements about their performance, certain fixed constructions may significantly outperform other constructions for small values of $B$. In fact, it is well-known that codebooks constructed by exploiting the channel correlation structure clearly outperform Grassmannian codebooks (and thus, in principle, RVQ codebooks) for small $B$ and that the condition number of the channel determines the performance of these codebooks [20], [25]–[29].

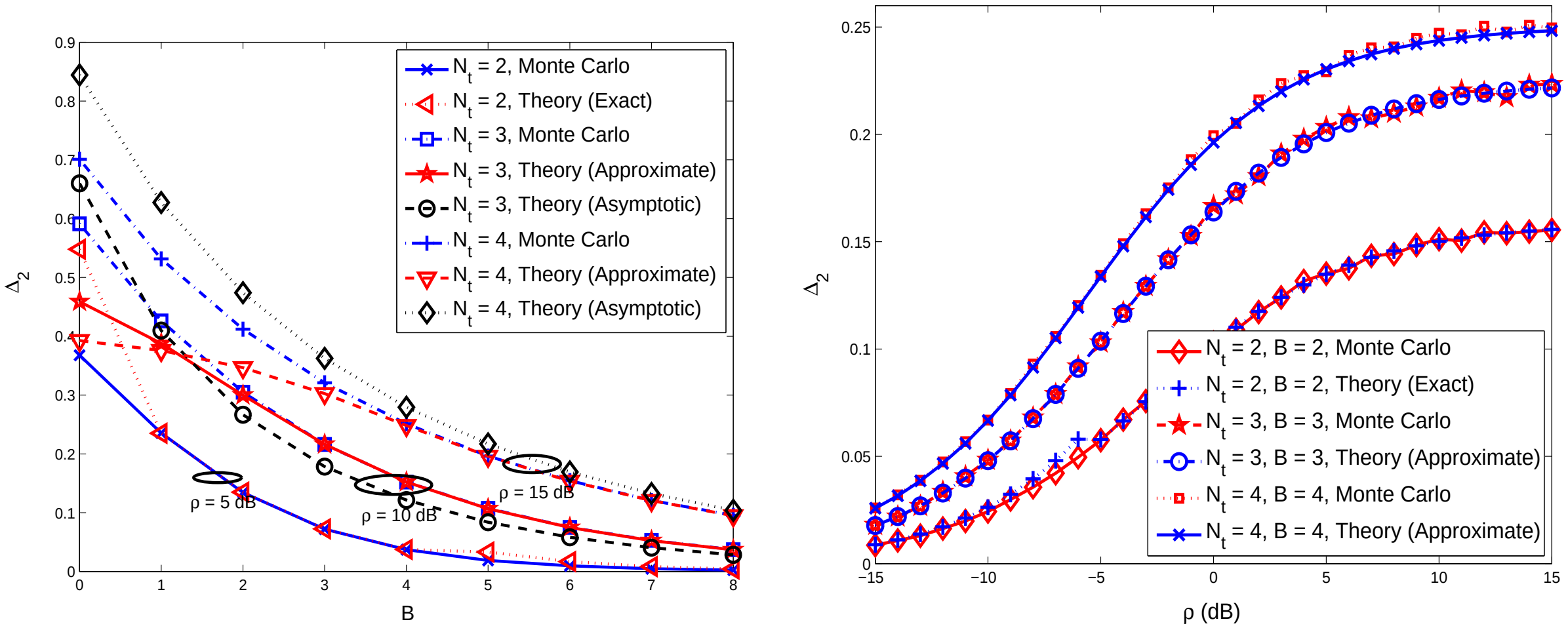

(a) (b)

Fig. 5. Instantaneous mutual information loss computed theoretically and via Monte Carlo averaging as a function of: a) $B$, and b) SNR.

To improve over the RVQ performance for finite values of $B$, we consider a codebook $\mathcal{C}_{\mathsf{sk}}$ where $\mathcal{C} = \left\{ \mathbf{f}_i,\, i = 1, \cdots, 2^B \right\}$ is skewed by a fixed $N_t \times N_t$ matrix $\mathsf{A}$ and then normalized as follows:

$$\mathcal{C}_{\mathsf{sk}} = \left\{ \frac{\mathsf{A}\,\mathbf{f}_i}{\|\mathsf{A}\,\mathbf{f}_i\|},\ i = 1, \cdots, 2^B \right\}. \tag{76}$$

The relative received SNR loss[3] with $\mathcal{C}_{\mathsf{sk}}$ is given as

$$\Delta_{1,\,\mathsf{sk}} = \mathbf{E}_{\mathcal{C}_{\mathsf{sk}}} \left[ 1 - \frac{1}{\lambda_1(\mathsf{H}^\dagger \mathsf{H})} \cdot \max_i \frac{\mathbf{f}_i^\dagger \mathsf{A}^\dagger \mathsf{H}^\dagger \mathsf{H} \mathsf{A} \mathbf{f}_i}{\mathbf{f}_i^\dagger \mathsf{A}^\dagger \mathsf{A} \mathbf{f}_i} \right] \tag{77}$$

and the broad goal is to design $\mathsf{A}_{\mathsf{opt}}$ where

$$\mathsf{A}_{\mathsf{opt}} = \arg \min_{\mathsf{A}} \mathbf{E}_{\mathbf{H}} \left[ \Delta_{1,\,\mathsf{sk}} \right]. \tag{78}$$

## A. Equivalent Characterization of $\Delta_{1,\,\mathsf{sk}}$

In this direction, a simple transformation argument $\mathbf{g}_i = \frac{\mathsf{A}\mathbf{f}_i}{\|\mathsf{A}\mathbf{f}_i\|} \in \mathcal{G}(N_t, 1)$ allows us to check that

$$\lambda_{N_t}(\mathsf{H}^\dagger \mathsf{H}) \leq \frac{\mathbf{f}_i^\dagger \mathsf{A}^\dagger \mathsf{H}^\dagger \mathsf{H} \mathsf{A} \mathbf{f}_i}{\mathbf{f}_i^\dagger \mathsf{A}^\dagger \mathsf{A} \mathbf{f}_i} \leq \lambda_1(\mathsf{H}^\dagger \mathsf{H}),\ i = 1, \cdots, 2^B. \tag{79}$$

[3]We will henceforth denote the explicit dependence of $\mathsf{H}^\dagger \mathsf{H}$ on $\lambda_1$ and use the notation $\lambda_1(\mathsf{H}^\dagger \mathsf{H})$ to distinguish it from the eigenvalues of $\mathsf{A}^\dagger \mathsf{A}$ and $\mathsf{A}^\dagger \mathsf{H}^\dagger \mathsf{H} \mathsf{A}$.

Further, along the lines of Lemma 1, it can also be checked that $\left\{\frac{\mathbf{f}_i^\dagger \mathsf{A}^\dagger \mathsf{H}^\dagger \mathsf{H} \mathsf{A} \mathbf{f}_i}{\mathbf{f}_i^\dagger \mathsf{A}^\dagger \mathsf{A} \mathbf{f}_i}, i = 1, \cdots, 2^B\right\}$ are i.i.d. Hence, as in Sec. III, we can rewrite $\Delta_{1,\mathsf{sk}}$ as

$$\Delta_{1,\mathsf{sk}} \cdot \lambda_1(\mathsf{H}^\dagger \mathsf{H}) = \int_{\lambda_{N_t}(\mathsf{H}^\dagger \mathsf{H})}^{\lambda_1(\mathsf{H}^\dagger \mathsf{H})} \left[ \Pr\left( \frac{\mathbf{f}^\dagger \mathsf{A}^\dagger \mathsf{H}^\dagger \mathsf{H} \mathsf{A} \mathbf{f}}{\mathbf{f}^\dagger \mathsf{A}^\dagger \mathsf{A} \mathbf{f}} \leq x \Big| \mathbf{f}^\dagger \mathbf{f} = 1 \right) \right]^m dx \tag{80}$$

where $\mathbf{f}$ is an isotropically distributed unit-norm random vector and $m = 2^B$. From (80), it is clear that quantifying $\Delta_{1,\mathsf{sk}}$ is dependent on knowledge of the distribution function of the ratio of weighted-norm of isotropically distributed unit-norm vectors. This is a hard problem, in general, unless there is some underlying structure to $\mathsf{A}$ that can be exploited. Of course, imposing structure on $\mathsf{A}$ cannot help solve for (78), an unconstrained optimization problem.

## B. Main Result

We overcome this technical difficulty by first studying the special case of $N_t = 2$. We then expand the intuition obtained from the $N_t = 2$ case to the more general case.

*Proposition 4:* In the special case of $N_t = 2$, $\Delta_{1,\mathsf{sk}}$ can be bounded by $\overline{\Delta}_{1,\mathsf{sk}}$, which behaves as $B \to \infty$ as:

$$\Delta_{1,\mathsf{sk}} \leq \overline{\Delta}_{1,\mathsf{sk}} \overset{B \to \infty}{\asymp} \frac{1}{2^B + 1} \cdot \left( 1 - \frac{\lambda_2(\mathsf{A}^\dagger \mathsf{H}^\dagger \mathsf{H} \mathsf{A})}{\lambda_1(\mathsf{H}^\dagger \mathsf{H}) \cdot \lambda_1(\mathsf{A} \mathsf{A}^\dagger)} \right). \tag{81}$$

*Proof:* The first step to prove the proposition is to establish a simplified version of (80). The second step deals with bounding $\Delta_{1,\mathsf{sk}}$ by an appropriate $\overline{\Delta}_{1,\mathsf{sk}}$ and capturing its asymptotic trend. See Appendix H for details. ■

Note that in the case of no skewing ($\mathsf{A} = \mathbf{I}$), (81) reduces to the result in Theorem 1.

*Theorem 4:* For the $N_t = 3$ case, the dominant term of an upper bound to $\Delta_{1,\mathsf{sk}}$ behaves as:

$$\Delta_{1,\mathsf{sk}} \cdot \lambda_1(\mathsf{H}^\dagger \mathsf{H}) \leq \overline{\Delta}_{1,\mathsf{sk}} \cdot \lambda_1(\mathsf{H}^\dagger \mathsf{H}) \tag{82}$$

$$= 2^{-\frac{B}{2}} \cdot \left[ 1 + \frac{\sqrt{\pi}}{2} \cdot \left( \frac{\lambda_1(\mathsf{A}^\dagger \mathsf{H}^\dagger \mathsf{H} \mathsf{A})}{\lambda_1(\mathsf{A}^\dagger \mathsf{A})} - \lambda_3(\mathsf{H}^\dagger \mathsf{H}) \right) \times \left( 1 + \frac{D_{\mathsf{sk}}}{(1 - D_{\mathsf{sk}})(N_t - 1)} \right) + G\left( \frac{\lambda_1(\mathsf{H}^\dagger \mathsf{H}) \lambda_1(\mathsf{A} \mathsf{A}^\dagger)}{\lambda_1(\mathsf{A}^\dagger \mathsf{H}^\dagger \mathsf{H} \mathsf{A})} \right) \right] + o\left(2^{-B/2}\right) \tag{83}$$

where

$$D_{\mathsf{sk}} = 1 - \prod_{j=2}^{N_t} \frac{\lambda_1(\mathsf{A}^\dagger \mathsf{H}^\dagger \mathsf{H} \mathsf{A}) - \lambda_{N_t}(\mathsf{H}^\dagger \mathsf{H}) \cdot \lambda_1(\mathsf{A}^\dagger \mathsf{A})}{\lambda_1(\mathsf{A}^\dagger \mathsf{H}^\dagger \mathsf{H} \mathsf{A}) - \lambda_j(\mathsf{A}^\dagger \mathsf{H}^\dagger \mathsf{H} \mathsf{A})} \tag{84}$$

and for some monotonically increasing function $G(\cdot)$, the structure of which is provided in (263)-(264) in Appendix I.

If $N_t \geq 4$, the asymptotic behavior (in $B$) of $\overline{\Delta}_{1,\,\mathsf{sk}}$ is as follows:

$$\Delta_{1,\,\mathsf{sk}} \leq \overline{\Delta}_{1,\,\mathsf{sk}} \tag{85}$$

$$= \underbrace{\frac{\kappa \cdot 2^{-\frac{B}{N_t-1}}}{N_t-1} \cdot \left(1 + \frac{D_{\mathsf{sk}}}{(1-D_{\mathsf{sk}})(N_t-1)}\right) \cdot \left(\frac{\lambda_1(\mathsf{A}^\dagger \mathsf{H}^\dagger \mathsf{H} \mathsf{A})}{\lambda_1(\mathsf{A}^\dagger \mathsf{A}) \cdot \lambda_1(\mathsf{H}^\dagger \mathsf{H})} - \frac{\lambda_{N_t}(\mathsf{H}^\dagger \mathsf{H})}{\lambda_1(\mathsf{H}^\dagger \mathsf{H})}\right)}_{\Delta_{1,\,\mathsf{sk},\,\mathsf{asymp}}}$$

$$+\, o\left(2^{-\frac{B}{N_t-1}}\right) \tag{86}$$

where $\kappa = \Gamma\left(\frac{1}{N_t-1}\right)$ and $D_{\mathsf{sk}}$ is as in (84).

*Proof:* See Appendix I. ■

### C. Insights on $\mathsf{A}_{\mathsf{opt}}$

While solving for $\mathsf{A}_{\mathsf{opt}}$ in (78) appears to be difficult, we now develop some insights on its structure.

1) From (1), recall that the system model (conditioned on $\mathbf{H} = \mathsf{H}$) with beamforming vector of index $i$ from $\mathcal{C}_{\mathsf{sk}}$ reduces to

$$\mathbf{y} = \sqrt{\frac{\rho}{\mathbf{f}_i^\dagger \mathsf{A}^\dagger \mathsf{A} \mathbf{f}_i}}\, \mathsf{H}\,\mathsf{A}\,\mathbf{f}_i\, s + \mathbf{n}. \tag{87}$$

By treating $\mathsf{HA}$ as the *effective channel* in (87), an application of Theorem 2 suggests that $\Delta_{1,\,\mathsf{sk}}$ is minimized if $\mathsf{HA}$ is well-conditioned. However, this argument is rigorous only if $\mathbf{f}_i^\dagger \mathsf{A}^\dagger \mathsf{A} \mathbf{f}_i$ can be treated as a constant for all $i$ so that $\mathsf{A}$ does not arbitrarily scale the power of the effective channel.

2) In the special case of $N_t = 2$, from Lemma 2, since $\mathbf{f}_i^\dagger \mathsf{A}^\dagger \mathsf{A} \mathbf{f}_i$ is uniformly distributed over the interval $\left[\lambda_2(\mathsf{A}^\dagger \mathsf{A}),\, \lambda_1(\mathsf{A}^\dagger \mathsf{A})\right]$, well-conditioning of $\mathsf{A}$ is necessary to ensure that $\mathbf{f}_i^\dagger \mathsf{A}^\dagger \mathsf{A} \mathbf{f}_i$ is approximately constant for all $\mathbf{f}_i$. Thus, there exists a tension between the two objectives (of well-conditioning of $\mathsf{HA}$ and $\mathsf{A}$) in deciding the appropriate choice of $\mathsf{A}$. Prop. 4 makes this intuition more concrete. From (81), it is clear that $\mathsf{A}$ should be chosen such that $\mathcal{L}_1$, defined as,

$$\mathcal{L}_1 \triangleq \frac{\lambda_1(\mathsf{A}\mathsf{A}^\dagger)}{\lambda_2(\mathsf{A}^\dagger \mathsf{H}^\dagger \mathsf{H} \mathsf{A})} \tag{88}$$

is minimized. But minimizing $\mathcal{L}_1$ is equivalent to minimizing the two squared condition numbers (of $\mathsf{HA}$ and $\mathsf{A}$), $\chi_{\mathsf{HA}} = \frac{\lambda_1(\mathsf{A}^\dagger \mathsf{H}^\dagger \mathsf{H} \mathsf{A})}{\lambda_2(\mathsf{A}^\dagger \mathsf{H}^\dagger \mathsf{H} \mathsf{A})}$ and $\chi_{\mathsf{A}} = \frac{\lambda_1(\mathsf{A}\mathsf{A}^\dagger)}{\lambda_2(\mathsf{A}\mathsf{A}^\dagger)}$. While a particular choice of $\mathsf{A}$ may make $\mathsf{HA}$ more well-conditioned than $\mathsf{H}$, this choice may not necessarily correspond to a well-conditioned $\mathsf{A}$ (and *vice versa*).

3) A further upper bound to the asymptotic trend in (83) of Theorem 4 (up to a multiplicative constant) in the $N_t = 3$ case is

$$\Delta_{1,\,\mathsf{sk}} \overset{\mathcal{O}(1)}{\leq} \frac{\left(1 - \frac{\lambda_2(\mathsf{A}^\dagger \mathsf{H}^\dagger \mathsf{H} \mathsf{A})}{\lambda_1(\mathsf{A}^\dagger \mathsf{H}^\dagger \mathsf{H} \mathsf{A})}\right) \cdot \left(1 - \frac{\lambda_3(\mathsf{A}^\dagger \mathsf{H}^\dagger \mathsf{H} \mathsf{A})}{\lambda_1(\mathsf{A}^\dagger \mathsf{H}^\dagger \mathsf{H} \mathsf{A})}\right)}{1 - \frac{\lambda_1(\mathsf{A}\mathsf{A}^\dagger)\lambda_3(\mathsf{H}^\dagger \mathsf{H})}{\lambda_1(\mathsf{A}^\dagger \mathsf{H}^\dagger \mathsf{H} \mathsf{A})}} + G\left(\frac{\lambda_1(\mathsf{H}^\dagger \mathsf{H}) \cdot \lambda_1(\mathsf{A}\mathsf{A}^\dagger)}{\lambda_1(\mathsf{A}^\dagger \mathsf{H}^\dagger \mathsf{H} \mathsf{A})}\right). \tag{89}$$

The goal of minimizing the term in (89) is equivalent to the goals of jointly minimizing

$$\mathcal{L}_2 \triangleq \frac{\lambda_1(\mathsf{A}^\dagger \mathsf{H}^\dagger \mathsf{H} \mathsf{A})}{\lambda_3(\mathsf{A}^\dagger \mathsf{H}^\dagger \mathsf{H} \mathsf{A})} \text{ and } \mathcal{L}_3 \triangleq \frac{\lambda_1(\mathsf{A}\mathsf{A}^\dagger)}{\lambda_1(\mathsf{A}^\dagger \mathsf{H}^\dagger \mathsf{H} \mathsf{A})}. \tag{90}$$

4) Consider the $N_t \geq 4$ case. Recasting Theorem 4, it can be seen that $\Delta_{1,\,\mathsf{sk},\,\mathsf{asymp}}$ is minimized if

$$\Big( N_t - 2 + \mathcal{L}_4 \Big) \cdot \mathcal{L}_5 \tag{91}$$

is also minimized, where

$$\mathcal{L}_4 \triangleq \prod_{j=2}^{N_t} \frac{\lambda_1(\mathsf{A}^\dagger \mathsf{H}^\dagger \mathsf{H} \mathsf{A}) - \lambda_j(\mathsf{A}^\dagger \mathsf{H}^\dagger \mathsf{H} \mathsf{A})}{\lambda_1(\mathsf{A}^\dagger \mathsf{H}^\dagger \mathsf{H} \mathsf{A}) - \lambda_{N_t}(\mathsf{H}^\dagger \mathsf{H}) \cdot \lambda_1(\mathsf{A}^\dagger \mathsf{A})} \tag{92}$$

$$\mathcal{L}_5 \triangleq \frac{\lambda_1(\mathsf{A}^\dagger \mathsf{H}^\dagger \mathsf{H} \mathsf{A}) - \lambda_{N_t}(\mathsf{H}^\dagger \mathsf{H}) \cdot \lambda_1(\mathsf{A}^\dagger \mathsf{A})}{\lambda_1(\mathsf{A}^\dagger \mathsf{A})}. \tag{93}$$

In the large-$N_t$ regime, observing that

$$\frac{\lambda_1(\mathsf{A}^\dagger \mathsf{H}^\dagger \mathsf{H} \mathsf{A}) - \lambda_j(\mathsf{A}^\dagger \mathsf{H}^\dagger \mathsf{H} \mathsf{A})}{\lambda_1(\mathsf{A}^\dagger \mathsf{H}^\dagger \mathsf{H} \mathsf{A}) - \lambda_{N_t}(\mathsf{H}^\dagger \mathsf{H}) \cdot \lambda_1(\mathsf{A}^\dagger \mathsf{A})} \geq 1 \tag{94}$$

for all $j$, we have

$$N_t - 2 \ll \mathcal{L}_4. \tag{95}$$

Thus, the dominant term of (91) in this regime is $\mathcal{L}_4 \cdot \mathcal{L}_5$.

5) Combining and unifying the above discussion, a (heuristically) "good" candidate for $\mathsf{A}$ should be such that the two metrics ($M_1$ and $M_2$), defined as,

$$M_1 \triangleq 1 - \frac{\lambda_1(\mathsf{A}^\dagger \mathsf{H}^\dagger \mathsf{H} \mathsf{A})}{\lambda_1(\mathsf{A}^\dagger \mathsf{A}) \cdot \lambda_1(\mathsf{H}^\dagger \mathsf{H})} \tag{96}$$

$$M_2 \triangleq \frac{\lambda_1(\mathsf{A}^\dagger \mathsf{H}^\dagger \mathsf{H} \mathsf{A})}{\lambda_{N_t}(\mathsf{A}^\dagger \mathsf{H}^\dagger \mathsf{H} \mathsf{A})} \tag{97}$$

are minimized jointly, if possible.

6) Conditioned on $\mathbf{H} = \mathsf{H}$, note that $M_1 \in [0,1]$ whereas $M_2 \in [1, \infty)$. The smallest value (of $0$) for $M_1$ is achieved with an $\mathsf{A}$ such that the eigenvectors of $\mathsf{A}\mathsf{A}^\dagger$ coincide with those of $\mathsf{H}^\dagger\mathsf{H}$ in the same order. With this choice, $M_2$ satisfies

$$M_2 = \chi_{\mathsf{H}} \cdot \chi_{\mathsf{A}}. \tag{98}$$

On the other hand, the smallest value (of $1$) for $M_2$ is achieved with $\mathsf{A} = \big(\mathsf{H}^\dagger\mathsf{H}\big)^{-1/2}$. With this choice, $M_1$ satisfies

$$M_1 = 1 - \frac{1}{\chi_{\mathsf{H}}}. \tag{99}$$

In other words, while $M_1$ is minimized by a choice of $\mathsf{A}$ whose left singular vectors *match* the right singular vectors of the channel, $M_2$ is minimized by a choice that *inverts*

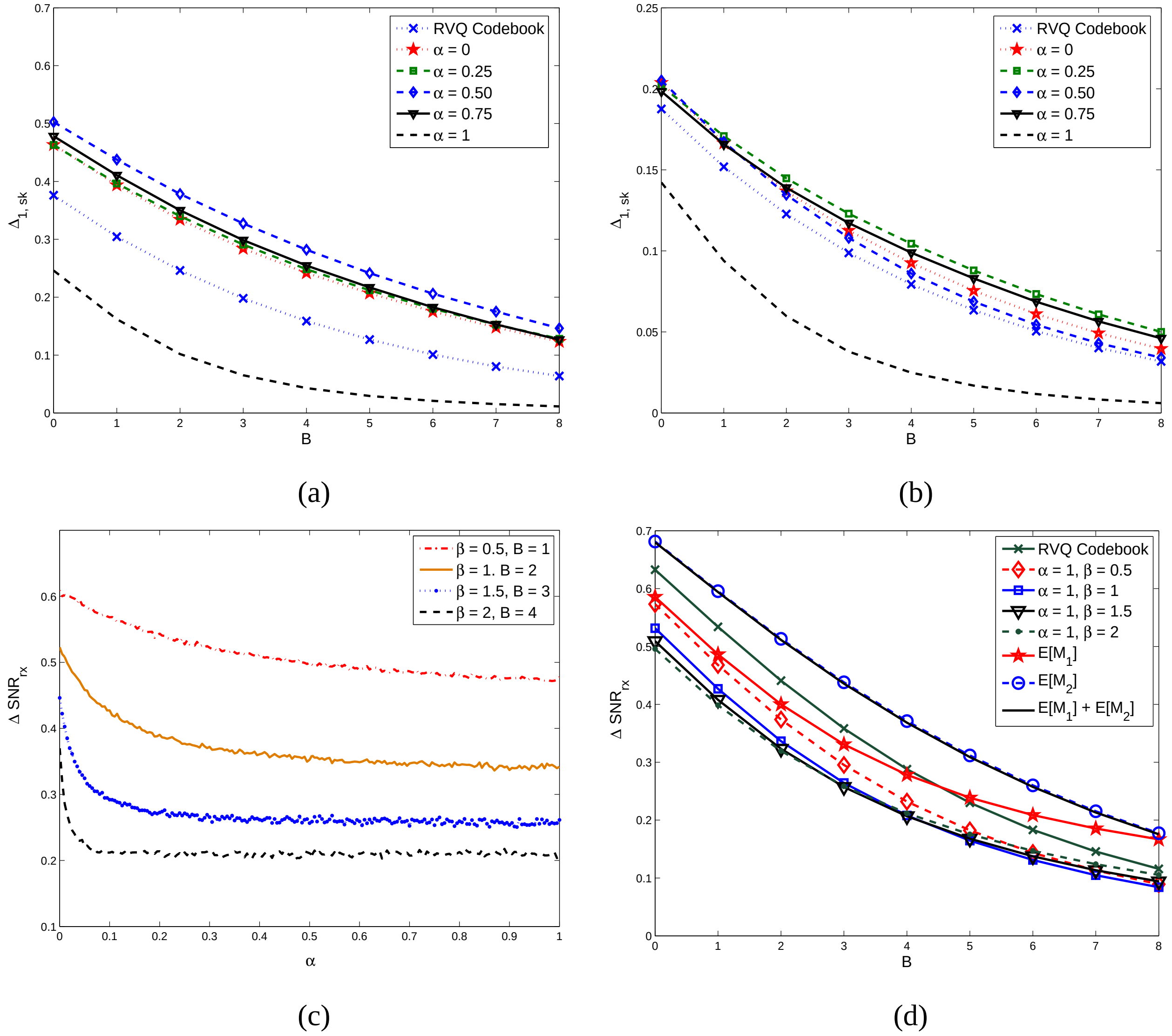

Fig. 6. Performance of skewed codebooks as a function of $B$ with a) an ill-conditioned channel realization, and b) a well-conditioned channel realization. Average performance of different families of skewed codebooks as a function of c) the parameter defining the skewing matrix classes, and d) $B$.

(or *zeroforces*) the channel. Thus, optimization over A is a combination of these two conflicting objectives in an appropriate sense. However, which of these objectives is more important than the other is not clear.

7) We now address this question via numerical studies for a fixed channel realization. In the first example, we consider an ill-conditioned channel with $N_t = N_r = 4$ and squared singular values $[4 \;\; 3 \;\; 2 \;\; 1]$. We numerically search over A to minimize

$$\mathcal{L}_6(\alpha) \triangleq \alpha \cdot M_1 + (1 - \alpha) \cdot M_2 \qquad (100)$$

for an appropriate choice of $\alpha \in [0, 1]$ that determines the weights between the two objectives in (96)-(97). The extreme cases of minimizing $M_1$ (or $M_2$) alone can be obtained by setting $\alpha = 1$ (or $\alpha = 0$) in (100). In Fig. 6(a), we plot the performance of the skewed codebooks (as a function of $B$) with $\mathsf{A}$ designed to minimize $\mathcal{L}_6(\alpha)$ for the following five choices of $\alpha$: i) $\alpha = 0$, ii) $\alpha = 0.25$, iii) $\alpha = 0.5$, iv) $\alpha = 0.75$, and v) $\alpha = 1$. The performance of the RVQ codebook ($\mathsf{A} = \mathbf{I}$) is also plotted. Fig. 6(a) shows that the goal of minimizing $M_1$ is more important than that of minimizing $M_2$ and the skewed codebook designed with this objective significantly out-performs the RVQ codebook (without skewing). In Fig. 6(b), we consider the performance of skewed codebooks designed for the same five choices of $\alpha$ (as above) in a well-conditioned channel with squared singular values given by $[1.6 \ 1.4 \ 1.2 \ 1]$. As in Fig. 6(a), we see that minimizing $M_1$ ($\alpha \approx 1$) is the more relevant objective in terms of limited-feedback performance. Figs. 6(a) and (b) also show that the performance with a poorly designed skewing matrix (e.g., $\alpha \approx 0$) can be significantly poorer than the RVQ performance.

8) With $\Delta\mathsf{SNR}_{\mathsf{rx}} = \mathbf{E}\left[\Delta_{1,\,\mathsf{sk}}\right]$ as the new metric, the previous study motivates the following family of matrices (parameterized by the weight $\alpha \in [0,1]$) for the design of skewed codebooks:

$$\mathsf{A}_{1,\,\alpha} = \arg\min_{\mathsf{A}} \ \mathbf{E}\left[\mathcal{L}_6(\alpha)\right] \tag{101}$$

$$= \arg\min_{\mathsf{A}} \ \alpha \cdot \mathbf{E}\left[M_1\right] + (1-\alpha) \cdot \mathbf{E}\left[M_2\right]. \tag{102}$$

While the family of skewing matrices within the argument in (101)-(102) is well-defined, a closed-form expression is hard to obtain for $\mathsf{A}_{1,\,\alpha}$.

To overcome this difficulty, consider the regime where $\{N_t, N_r\} \to \infty$ with $\frac{N_t}{N_r} \to 0$. Using the channel hardening principle in this regime [21], [46], the eigenvectors of $\mathsf{A}\mathsf{A}^\dagger$ for the choice of $\mathsf{A}$ that minimizes $\mathbf{E}\left[M_1\right]$ can be heuristically replaced with the eigenvectors of $\mathbf{E}\left[\mathsf{H}^\dagger\mathsf{H}\right] = \mathbf{\Sigma}_t$. A suitable candidate[4] for such an $\mathsf{A}$ is

$$\mathsf{A} = \left(\mathbf{\Sigma}_t\right)^{\beta} = \mathbf{U}_t \left(\mathbf{\Lambda}_t\right)^{\beta} \mathbf{U}_t^\dagger \tag{103}$$

for some choice of $\beta$ satisfying $\beta \geq 0$. Similarly, the choice of $\mathsf{A}$ that minimizes $\mathbf{E}\left[M_2\right]$ can be heuristically replaced by

$$\mathsf{A} = \left(\mathbf{\Sigma}_t\right)^{-\frac{1}{2}} = \mathbf{U}_t \left(\mathbf{\Lambda}_t\right)^{-\frac{1}{2}} \mathbf{U}_t^\dagger. \tag{104}$$

We interpolate the two statistics-dependent candidates in (103) and (104) to obtain the following family of matrices for skewing:

$$\mathsf{A}_{2,\,\alpha,\,\beta} = \alpha \cdot \left(\mathbf{\Sigma}_t\right)^{\beta} + (1-\alpha) \cdot \left(\mathbf{\Sigma}_t\right)^{-\frac{1}{2}} \tag{105}$$

$$= \mathbf{U}_t \left(\alpha(\mathbf{\Lambda}_t)^{\beta} + (1-\alpha)(\mathbf{\Lambda}_t)^{-\frac{1}{2}}\right) \mathbf{U}_t^\dagger \tag{106}$$

[4]Note that since the eigenvectors of $\mathsf{A}\mathsf{A}^\dagger$ have to be in the same order as (the order of) the eigenvectors of $\mathsf{H}^\dagger\mathsf{H}$ to minimize $M_1$, this constraint can only be ensured by setting $\beta \geq 0$ in (103).

for some $\alpha \in [0,1]$ and $\beta \geq 0$. Note that the right-hand side of (106) can also be written as

$$\mathsf{A}_{2,\,\alpha,\,\beta} = \mathbf{U}_t \, h\left(\mathbf{\Lambda}_t\right) \, \mathbf{U}_t^{\dagger} = h\left(\mathbf{\Sigma}_t\right) \tag{107}$$

for an appropriate choice of the matrix function $h(\cdot)$. In this sense, (107) generalizes the skewing matrix proposed in [25] (which can be obtained by setting $\alpha = 1, \beta = \frac{1}{2}$) and the matrix proposed in [26] (which can be obtained by setting $\alpha = 1 = \beta$).

9) We now numerically study the $\Delta\mathsf{SNR}_{\mathsf{rx}}$ performance of codebooks obtained by skewing an RVQ codebook with the two families: $\{\mathsf{A}_{1,\,\alpha}\}$ and $\{\mathsf{A}_{2,\,\alpha,\,\beta}\}$. We consider a Kronecker-product correlated channel with $\mathbf{\Sigma}_t = 1.6 \times \mathsf{diag}([4 \;\; 3 \;\; 2 \;\; 1])$ and $\mathbf{\Sigma}_r = \mathsf{diag}([7 \;\; 5 \;\; 3 \;\; 1])$. Note that the channel power is normalized as $\mathsf{Tr}(\mathbf{\Sigma}_t) = \mathsf{Tr}(\mathbf{\Sigma}_r) = 16$.

   In the first study, we plot $\Delta\mathsf{SNR}_{\mathsf{rx}}$ as a function of $\alpha$ for the $\{\mathsf{A}_{2,\,\alpha,\,\beta}\}$ family for different values of $\beta$ and $B$ in Fig. 6(c). For all the $\{\beta, B\}$ combinations considered, the smallest value of $\Delta\mathsf{SNR}_{\mathsf{rx}}$ is achieved as $\alpha \to 1$, thereby justifying the following study where attention is restricted to the case of $\alpha = 1$ from the $\{\mathsf{A}_{2,\,\alpha,\,\beta}\}$ family.

10) In the second study, a numerical search over $\mathsf{A}$ is performed with the objective of minimizing: i) $\mathbf{E}[M_1]$, ii) $\mathbf{E}[M_1] + \mathbf{E}[M_2]$, and iii) $\mathbf{E}[M_2]$, corresponding to three choices from $\{\mathsf{A}_{1,\,\alpha}\}$: i) $\mathsf{A}_{1,\,\alpha=1}$, ii) $\mathsf{A}_{1,\,\alpha=0.5}$, and iii) $\mathsf{A}_{1,\,\alpha=0}$, respectively. Motivated by the study in Fig. 6(c), four other skewing matrices from the $\{\mathsf{A}_{2,\,\alpha=1,\,\beta}\}$ family are also considered: iv) $\mathsf{A}_{2,\,\alpha=1,\,\beta=0.5}$, v) $\mathsf{A}_{2,\,\alpha=1,\,\beta=1}$, vi) $\mathsf{A}_{2,\,\alpha=1,\,\beta=1.5}$, and vii) $\mathsf{A}_{2,\,\alpha=1,\,\beta=2}$. Note that as stated previously, iv) and v) correspond to the skewing matrix choices proposed in [25] and [26], respectively.

    Fig. 6(d) plots $\Delta\mathsf{SNR}_{\mathsf{rx}}$ (as a function of $B$) for these seven skewed codebooks as well as the RVQ codebook and we see that $\mathsf{A}_{1,\,\alpha=1}$ results in better performance over RVQ codebooks for small $B$ values ($B \leq 4$). However, as $B$ increases, the average performance with this choice of skewing matrix deteriorates over an RVQ codebook. On the other hand, both $\mathsf{A}_{1,\,\alpha=0.5}$ as well as $\mathsf{A}_{1,\,\alpha=0}$ result in poorer performance relative to the RVQ scheme thus confirming the importance of $M_1$ over $M_2$ in skewing matrix optimization. We also see that the $\{\mathsf{A}_{2,\,\alpha=1,\,\beta}\}$ family results in improved performance over the $\{\mathsf{A}_{1,\,\alpha}\}$ family as well as RVQ codebooks. Further, the performance with skewing matrices for values of $\beta$ satisfying $\beta > 1$ from the $\{\mathsf{A}_{2,\,\alpha=1,\,\beta}\}$ family is better than that achieved with the choices $\beta = 0.5$ and $\beta = 1$.

    In general, we observe that for fixed $\beta$ values, as $\alpha$ increases, performance gets better for any $B$, with the performance becoming independent of $\alpha$ for large values of $\beta$. For fixed $\alpha$, large $\beta$ is seen to be better for small $B$ values ($B \approx 0-3$) whereas $\beta = 1$ is robust for large $B$ values ($B \approx 7-8$). Similar behavior is observed with other choices of transmit and receive covariance matrices furnishing evidence to the observations in the literature that appropriately designed skewed codebooks can significantly out-perform RVQ codebooks over correlated channels.

## VII. Conclusion

Limited-feedback communications has become an important component of 3G/4G cellular standardization efforts. However, performance analysis of limited-feedback schemes, especially under practical impairments such as channel correlation, has not received much attention in the literature. The main goal of this work is to study the ensemble properties of a $B$-bit RVQ codebook in the correlated MIMO setting with the metrics of interest being the received $\mathsf{SNR}$ loss ($\Delta\mathsf{SNR}_{\mathsf{rx}}$) and loss in average mutual information ($\Delta I$), both relative to a perfect CSI scheme.

We computed the rate of decay of $\Delta\mathsf{SNR}_{\mathsf{rx}}$ and $\Delta I$ as a function of $B$ and the channel correlation profile. While the rate of decay with $B$ is in conformance with similar results obtained in the literature for i.i.d. MIMO/MISO/$\mathsf{rank}$-1 MIMO channels [12], [15], [30]–[37], our result applies to correlated MIMO channels of arbitrary rank and arbitrary choice of $B$. For fixed $B$, the critical factor limiting the RVQ performance is the condition number of the channel. We established that the channel correlation profile that minimizes the performance loss with an RVQ codebook is typically i.i.d.-like (spatially rich) and the profile that maximizes the performance loss has $\mathsf{rank}$-1 (spatially poor/sparse structure). This result on the dependence of RVQ performance on the condition number should not be entirely surprising [20], [21], [31] given that the RVQ codebook consists of isotropic beamforming vectors and an i.i.d. channel has dominant right singular vector that is also isotropic.

We then generalized our performance analysis to the case of skewed codebooks where the RVQ codebook is skewed by a fixed matrix and normalized to ensure unit-norm. From this characterization, we showed that the tension between well-conditioning of the *effective channel* and well-conditioning of the skewing matrix determines the structure of the optimal skewing matrix for limited-feedback beamforming. In particular, we established the criticality of matching between the left singular vectors of the skewing matrix and the right singular vectors of the channel. Using this insight, we constructed a class of statistics-dependent (more specifically, transmit covariance matrix-dependent) skewing matrices that result in significantly improved performance over RVQ codebooks.

The workhorse behind our study is the structure of the density function of weighted-norm of isotropically distributed unit-norm vectors. This tool plays an important role in other settings such as precoder design for broadcast [47] and interference channels [48], and norm feedback in broadcast channels [17]. Notwithstanding the results of this paper, the characterization of the performance loss with skewed codebooks is incomplete. Generalizing our toolkit to the density function of the ratio of weighted-norms is important in establishing fundamental performance limits with skewed codebooks (which are linear by definition) as well as non-linear skewed codebooks as constructed in [20], moment and distributional properties on the various performance metrics, identifying the structure of the optimal skewing matrix, etc. Other problems of interest in the single-user setting include averaging the loss expressions over the channel randomness to study the impact of the channel model (Kronecker vs. non-Kronecker) on performance, establishing possible majorization results for performance metrics as a function of the transmit and receive covariance matrix eigenvalues, performance of higher-rank schemes [34], [49], etc. Extension of this study to the multi-user setting [15]–[19] is also of practical interest.

## Appendix

### A. Proof of Lemma 1

For the first statement, for any $\Theta = [\Theta_1, \cdots, \Theta_{N_t}]$, note from (8) that

$$\Pr\Big(|\mathbf{f}_i(\pi_1)|^2, \cdots, |\mathbf{f}_i(\pi_{N_t})|^2 \in \Theta\Big) = \int_{\mathbf{f}_i \,:\, \{|\mathbf{f}_i(\pi_k)|^2 \in \Theta_k\}} \frac{\Gamma(N_t)}{\pi^{N_t}} \cdot \delta\big(\mathbf{f}_i^\dagger \mathbf{f}_i - 1\big)\, d\mathbf{f}_i \tag{108}$$

$$= \frac{\Gamma(N_t)}{\pi^{N_t}} \cdot \mathsf{Area}\big(\mathbf{f}_i \,:\, \{|\mathbf{f}_i(\pi_k)|^2 \in \Theta_k\} | \mathbf{f}_i^\dagger \mathbf{f}_i = 1\big). \tag{109}$$

Since $\mathbf{f}_i$ is isotropic on $\mathcal{G}(N_t, 1)$, (109) is circularly symmetric and hence, independent of the permutation $\Pi$.

For the second statement, the Ritz-Rayleigh relationship implies that the range of $\mathbf{x}_i$ is $[\lambda_{N_t}, \lambda_1]$. The independence of $\{\mathbf{x}_i, i = 1, \cdots, 2^B\}$ follows from the independence of $\{\mathbf{f}_i, i = 1, \cdots, 2^B\}$. To prove that $\{\mathbf{x}_i\}$ are also identically distributed, note that if $\{\mathbf{f}_i\}$ are isotropic and i.i.d., then so are $\{\mathbf{g}_i = \mathsf{U}^\dagger \mathbf{f}_i\}$ for any fixed unitary matrix $\mathsf{U}$. The fixed unitary matrix in this setting is the eigenvector matrix in an eigen-decomposition of $\mathsf{H}^\dagger\mathsf{H}$ for a given realization $\mathsf{H}$, wherein we have $\mathsf{H}^\dagger\mathsf{H} = \mathsf{U}\Lambda\mathsf{U}^\dagger$. The diagonal matrix $\Lambda = \mathsf{diag}\left([\lambda_1, \cdots, \lambda_{N_t}]\right)$ is in general not the identity matrix. For any fixed $k$, $\{|\mathbf{g}_i(k)|^2, \ i = 1, \cdots, 2^B\}$ are identically distributed since $\{\mathbf{g}_i\}$ are i.i.d. The conclusion follows since $\mathbf{x}_i = \sum_k |\mathbf{g}_i(k)|^2 \lambda_k$. ■

### B. Proof of Lemma 2

Following the derivation of the density function of $\mathbf{f}^\dagger\Lambda\mathbf{f}$ when $\lambda_2 = \cdots = \lambda_{N_t} = 0$ in [8], we have

$$\mathsf{P}(x) \;\triangleq\; \Pr\left(\mathbf{f}^\dagger\Lambda\mathbf{f} = x\right) = \frac{\partial}{\partial x}\Pr\left(\mathbf{f}^\dagger\Lambda\mathbf{f} \leq x\right) \tag{110}$$

with

$$\Pr\left(\mathbf{f}^\dagger\Lambda\mathbf{f} \leq x\right) \;=\; 1 - \frac{\mathsf{Area}\,(x,\,1)}{\mathsf{Area}\,(1)} \tag{111}$$

where

$$\mathsf{Area}\,(x,\,y) \triangleq \mathsf{Area}\,\left(\mathbf{f}^\dagger\Lambda\mathbf{f} \geq x,\, \|\mathbf{f}\|^2 = y\right) \text{ and} \tag{112}$$

$$\mathsf{Area}\,(y) \triangleq \mathsf{Area}\,\left(\|\mathbf{f}\|^2 = y\right) \tag{113}$$

denote the area of a (unit-radius) sphere carved out by the ellipsoid $\{\mathbf{f} : \mathbf{f}^\dagger \Lambda \mathbf{f} = x\}$ and the area of a (unit-radius) complex sphere, respectively. The volume of the objects desired in the computation of $\mathsf{P}(x)$ are

$$\begin{aligned} \mathsf{Vol}\left(x, r^2\right) &\triangleq \mathsf{Vol}\left(\mathbf{f}^\dagger \Lambda \mathbf{f} \geq x,\ \|\mathbf{f}\|^2 \leq r^2\right) \quad (114)\\ &= \int_{y=0}^{r^2} \mathsf{Area}\,(x,y)\, dy \ \ \text{and} \quad (115) \end{aligned}$$

$$\mathsf{Vol}(r^2) \triangleq \mathsf{Vol}\left(\|\mathbf{f}\|^2 \leq r^2\right) = \int_{x=0}^{r^2} \mathsf{Area}(x) dx. \quad (116)$$

Thus, we have

$$\begin{aligned} \mathsf{Area}\,(x,1) &= \frac{\partial}{\partial r^2} \mathsf{Vol}\left(x, r^2\right)\Big|_{r=1}, \quad (117)\\ \mathsf{Area}\,(1) &= \frac{\partial}{\partial r^2} \mathsf{Vol}(r^2)\Big|_{r=1} \ \text{and hence,} \quad (118)\\ \mathsf{P}(x) &= -\frac{\frac{\partial^2}{\partial x r^2} \mathsf{Vol}\,(x, r^2)\Big|_{r=1}}{\frac{\partial}{\partial r^2} \mathsf{Vol}\,(r^2)\Big|_{r=1}}. \quad (119) \end{aligned}$$

Computing $\mathsf{Vol}\,(x, r^2)$ is non-trivial even in the simple case of $N_t = 2$. This is because every additional dimension to the complex ellipsoid corresponds to addition of two real dimensions. In the simplest case of $N_t = 2$, we have the intersection of two four-dimensional real objects which cannot be visualized pictorially. Nevertheless, the following lemma captures the complete structure of $\mathsf{P}(x)$ when $N_t = 2$. The general case follows subsequently.

*Lemma 3:* If $N_t = 2$, the random variable $\mathbf{f}^\dagger \Lambda \mathbf{f}$ is uniformly distributed in the interval $[\lambda_2, \lambda_1]$. ■

*Proof:* First, note that it follows from [8, Lemma 2] that

$$\mathsf{Vol}(r^2) = \frac{\pi^{N_t} r^{2N_t}}{\Gamma(N_t + 1)}. \quad (120)$$

For computing $\mathsf{Vol}\,(x, r^2)$, we follow the same variable transformation as in [8]. We set $\mathbf{f}(k) = r_k \exp(j\theta_k)$ for $k = 1, 2$. The ellipsoid is contained completely in the sphere of radius $r$ if $r$ is such that $r \geq \sqrt{\frac{x}{\lambda_2}}$ whereas the sphere is contained completely in the ellipsoid if $r \leq \sqrt{\frac{x}{\lambda_1}}$. In the intermediate regime for $r$, a non-trivial intersection between the two objects is observed and one can compute the volume by performing a two-dimensional integration as follows:

$$\begin{aligned} \mathsf{Vol}\left(x, r^2\right) &= \iint_{\mathcal{A}} r_1 r_2 dr_1 dr_2 d\theta_1 d\theta_2 \quad (121)\\ &= (2\pi)^2 \cdot \iint_{\mathcal{B}} r_1 dr_1 r_2 dr_2 \quad (122)\\ &= (2\pi)^2 \cdot \int_0^{r^\star} r_2 dr_2 \cdot \int_{L'}^{U'} r_1 dr_1 \quad (123) \end{aligned}$$

where

$$\mathcal{A} = \left\{r_1, r_2 \ : \ r_1^2\lambda_1 + r_2^2\lambda_2 \geq x, \ r_1^2 + r_2^2 \leq r^2\right\} \text{ and } \{\theta_1, \theta_2 \ : \ [0, 2\pi)\} \tag{124}$$

$$\mathcal{B} = \left\{r_1, r_2 \ : \ r_1^2\lambda_1 + r_2^2\lambda_2 \geq x, \ r_1^2 + r_2^2 \leq r^2\right\} \tag{125}$$

$$L' = \sqrt{\frac{x - r_2^2\,\lambda_2}{\lambda_1}} \tag{126}$$

$$U' = \sqrt{r^2 - r_2^2} \tag{127}$$

$$r^\star = \frac{r^2\,\lambda_1 - x}{\lambda_1 - \lambda_2}. \tag{128}$$

Straight-forward computation from (123) establishes the following:

$$\mathsf{Vol}\left(x, r^2\right) = \begin{cases} 0 & r \leq \sqrt{\frac{x}{\lambda_1}} \\ \frac{\pi^2}{2} \cdot \frac{\left(r^2\,\lambda_1 - x\right)^2}{\lambda_1(\lambda_1 - \lambda_2)} & \sqrt{\frac{x}{\lambda_1}} \leq r \leq \sqrt{\frac{x}{\lambda_2}} \\ \frac{\pi^2}{2} \cdot \left(r^4 - \frac{x^2}{\lambda_1\lambda_2}\right) & r \geq \sqrt{\frac{x}{\lambda_2}} \end{cases} \tag{129}$$

Using (119), another trivial computation shows that

$$\mathsf{P}(x) = \frac{1}{\lambda_1 - \lambda_2}, \ \lambda_2 \leq x \leq \lambda_1. \tag{130}$$

That is, $\mathbf{f}^\dagger \Lambda \mathbf{f}$ is uniformly distributed in its range. ■

*Lemma 4:* This lemma states (without proof) the structure of the density function $\mathsf{P}(x)$ in the cases $N_t = 3$ and $N_t = 4$. With $N_t = 3$, we have

$$\mathsf{P}(x) = \begin{cases} 0 & x \leq \lambda_3 \\ \frac{2\,(x - \lambda_3)}{(\lambda_1 - \lambda_3)(\lambda_2 - \lambda_3)} & \lambda_3 \leq x \leq \lambda_2 \\ \frac{2\,(\lambda_1 - x)}{(\lambda_1 - \lambda_2)(\lambda_1 - \lambda_3)} & \lambda_2 \leq x \leq \lambda_1 \\ 0 & x \geq \lambda_1. \end{cases} \tag{131}$$

With $N_t = 4$, we have

$$\mathsf{P}(x) = \begin{cases} 0 & x \leq \lambda_4 \\ \frac{3\,(x - \lambda_4)^2}{(\lambda_1 - \lambda_4)(\lambda_2 - \lambda_4)(\lambda_3 - \lambda_4)} & \lambda_4 \leq x \leq \lambda_3 \\ \kappa_1 & \lambda_3 \leq x \leq \lambda_2 \\ \frac{3\,(\lambda_1 - x)^2}{(\lambda_1 - \lambda_2)(\lambda_1 - \lambda_3)(\lambda_1 - \lambda_4)} & \lambda_2 \leq x \leq \lambda_1 \\ 0 & x \geq \lambda_1 \end{cases} \tag{132}$$

where

$$\kappa_1 = \frac{3}{(\lambda_1 - \lambda_3)\,(\lambda_2 - \lambda_4)} \cdot \kappa_2 \tag{133}$$

$$\kappa_2 = \frac{(x - \lambda_3)\,(\lambda_2 - x)}{\lambda_2 - \lambda_3} + \frac{(x - \lambda_4)\,(\lambda_1 - x)}{\lambda_1 - \lambda_4}. \tag{134}$$

■

*C. Proof of Theorem 1*

As stated at the beginning of Sec. III, we compute $\Delta_1$ using Lemma 2. The computation of $\Delta_1$ in the $N_t = 2$ case is a straight-forward integration.

For the $N_t = 3$ case, we split the integral computation into two parts: the intervals $[\lambda_3, \lambda_2]$ and $[\lambda_2, \lambda_1]$. The integral over the first interval is again straight-forward and results in the contribution of

$$\frac{1}{2m+1} \cdot \frac{\lambda_2 - \lambda_3}{\lambda_1} \cdot \left( \frac{\lambda_2 - \lambda_3}{\lambda_1 - \lambda_3} \right)^m . \tag{135}$$

Upon elementary manipulation, the integral over the second interval can be shown to be equivalent to

$$\frac{\sqrt{(\lambda_1 - \lambda_2)(\lambda_1 - \lambda_3)}}{\lambda_1} \cdot \int_{y=0}^{\sqrt{\frac{\lambda_1 - \lambda_2}{\lambda_1 - \lambda_3}}} \left(1 - y^2\right)^m dy \tag{136}$$

which can be computed in closed-form using integral tables [50, 2.512(3), p. 131] via the transformation $y \mapsto \sin(\theta)$. Combining the two terms, we have the expression for $\Delta_1$ in the statement of the theorem.

For the $N_t \geq 4$ case, exact computation of $\Delta_1$ is cumbersome. Since the distribution function $F(x)$ is monotonically increasing, the dominant trend (and term) of $\Delta_1$ is captured by the integral over the segment $[\lambda_2, \lambda_1]$ alone. This integral can be computed in closed-form due to the tractable nature of $F(x)$ in this interval. Upon elementary transformations, this integral is seen to be:

$$\mathsf{C}_2 \cdot \int_{\theta=0}^{\theta_{\max}} \cos^{2m+1}(\theta) \sin^p(\theta) \; d\theta \tag{137}$$

with $p = \frac{2}{N_t - 1} - 1$,

$$\mathsf{C}_2 = \frac{2}{(N_t - 1)} \cdot \left[ \prod_{j=2}^{N_t} \left( 1 - \frac{\lambda_j}{\lambda_1} \right) \right]^{\frac{1}{N_t - 1}} \tag{138}$$

$$\theta_{\max} = \sin^{-1} \left( \sqrt{ \prod_{j=2}^{N_t} \frac{\lambda_1 - \lambda_2}{\lambda_1 - \lambda_j} } \right) . \tag{139}$$

Again, using the integral tables [50, 2.511(4), p. 131], we can compute (137) in closed-form as in the statement of the theorem.

It is obvious that $\Delta_{1,\,\mathsf{appx}} \le \Delta_1$. For the other side of (30), note that

$$1 - \frac{\Delta_{1,\,\mathsf{appx}}}{\Delta_1} = \frac{\int_{\lambda_{N_t}}^{\lambda_2} \left(\Pr\left(\mathbf{f}^\dagger \Lambda \mathbf{f} \le x\right)\right)^m dx}{\int_{\lambda_{N_t}}^{\lambda_1} \left(\Pr\left(\mathbf{f}^\dagger \Lambda \mathbf{f} \le x\right)\right)^m dx} \tag{140}$$

$$\le \frac{\frac{1}{\lambda_1} \int_{\lambda_{N_t}}^{\lambda_2} \left(\Pr\left(\mathbf{f}^\dagger \Lambda \mathbf{f} \le x\right)\right)^m dx}{\Delta_{1,\,\mathsf{appx}}} \tag{141}$$

$$\le \frac{\left(\Pr\left(\mathbf{f}^\dagger \Lambda \mathbf{f} \le \lambda_2\right)\right)^m (\lambda_2 - \lambda_{N_t})}{\lambda_1 \cdot \Delta_{1,\,\mathsf{appx}}} \tag{142}$$

$$= \frac{\lambda_2 - \lambda_{N_t}}{\lambda_1} \cdot \frac{D^m}{\Delta_{1,\,\mathsf{appx}}} \tag{143}$$

where the third step follows by bounding the distribution by its largest value at $x = \lambda_2$ and the last step by noting from (25) that

$$\Pr\left(\mathbf{f}^\dagger \Lambda \mathbf{f} \le \lambda_2\right) = D. \tag{144}$$

■

*D. Proof of Prop. 1*

In the general MIMO setting with $N_t = 3$, we have

$$\frac{2^{k+1} m(m-1)\cdots(m-k)}{(2m-1)(2m-3)\cdots(2m-2k-1)} = \frac{h(m)}{h(m-k-1)} \tag{145}$$

where $h(\cdot)$ is a function defined on the set of integers as

$$h(m) \triangleq \frac{(m!)^2 \cdot 2^{2m}}{2m!}. \tag{146}$$

Using Stirling's formula [51, 6.1.39, p. 257] to approximate the factorial function as $m = 2^B$ increases, we obtain a good estimate of the trend of $h(m)$, and hence the summation in the characterization of $\Delta_1$ in Theorem 1. Retaining the dominant terms, we can write $\Delta_1$ as

$$\Delta_1 \overset{B\to\infty}{\asymp} \frac{\sqrt{\pi}}{2^{B/2+1}} \cdot \left[\left(1 - \frac{\lambda_2}{\lambda_1}\right)\left(1 + \frac{\lambda_2 - \lambda_3}{2(\lambda_1 - \lambda_3)}\right)\right]. \tag{147}$$

In the $N_t \ge 4$ case, we have

$$\frac{2^k \cdot m(m-1)\cdots(m-k+1)}{\prod_{j=m-k}^{m-1}(p+1+2j)} = \frac{m! \cdot \Gamma\left(\frac{1}{N_t-1} + m - k\right)}{\Gamma\left(\frac{1}{N_t-1} + m\right) \cdot m-k!} \tag{148}$$

where $\Gamma(\cdot)$ stands for the Gamma function. With $k = m$, the above equation simplifies to

$$\frac{m! \cdot \Gamma\left(\frac{1}{N_t-1}\right)}{\Gamma\left(\frac{1}{N_t-1} + m\right)} \overset{B\to\infty}{\asymp} \frac{\sqrt{2\pi m} \cdot m^m \cdot e^{-m} \cdot \Gamma\left(\frac{1}{N_t-1}\right)}{\sqrt{2\pi} \cdot e^{-m} \cdot m^{m+\frac{1}{N_t-1}-\frac{1}{2}}} \tag{149}$$

$$= m^{1-\frac{1}{N_t-1}} \cdot \Gamma\left(\frac{1}{N_t-1}\right) \tag{150}$$

where the asymptotic trend follows from Stirling's formula. For $1 \leq k \leq m-1$, we have

$$\frac{m! \cdot \Gamma\left(\frac{1}{N_t-1}+m-k\right)}{\Gamma\left(\frac{1}{N_t-1}+m\right) \cdot m-k!} \stackrel{B\to\infty}{\asymp} m^{1-\frac{1}{N_t-1}} \cdot \frac{\Gamma\left(m-k+\frac{1}{N_t-1}\right)}{\Gamma\left(m-k+1\right)}. \tag{151}$$

Using the trivial inequality $\frac{\Gamma\left(m-k+\frac{1}{N_t-1}\right)}{\Gamma(m-k+1)} \leq \frac{\kappa}{N_t-1}$ with $\kappa = \Gamma\left(\frac{1}{N_t-1}\right)$, we have

$$\Delta_1 \leq \frac{\kappa \cdot m^{1-\frac{1}{N_t-1}}}{A_{N_t}} \cdot \left[\left(1-\frac{\lambda_2}{\lambda_1}\right) \cdot \left(1+\frac{D}{(1-D)(N_t-1)}\right)\right] \tag{152}$$

$$\asymp \frac{\kappa \cdot 2^{-\frac{B}{N_t-1}}}{N_t-1} \cdot \left[\left(1-\frac{\lambda_2}{\lambda_1}\right) \cdot \left(1+\frac{D}{(1-D)(N_t-1)}\right)\right]. \tag{153}$$

■

*E. Proof of Theorem 2*

First, note from (15) that (47) is equivalent to showing that

$$\frac{\mathbf{E}_{\mathcal{C}}\left[\max_i \mathbf{f}_i^{\dagger} \mathsf{H}_1^{\dagger} \mathsf{H}_1 \mathbf{f}_i\right]}{\lambda_1} \geq \frac{\mathbf{E}_{\mathcal{C}}\left[\max_i \mathbf{f}_i^{\dagger} \mathsf{H}_2^{\dagger} \mathsf{H}_2 \mathbf{f}_i\right]}{\mu_1}. \tag{154}$$

Using the eigen-decompositions

$$\mathsf{H}_1^{\dagger}\mathsf{H}_1 = \mathsf{U}_1 \,\mathsf{diag}(\underline{\boldsymbol{\lambda}})\, \mathsf{U}_1^{\dagger}, \quad \mathsf{H}_2^{\dagger}\mathsf{H}_2 = \mathsf{U}_2 \,\mathsf{diag}(\underline{\boldsymbol{\mu}})\, \mathsf{U}_2^{\dagger} \tag{155}$$

in (154), we have

$$\frac{\mathbf{E}_{\mathcal{C}}\left[\max_i \mathbf{f}_i^{\dagger}\, \mathsf{U}_1 \,\mathsf{diag}(\underline{\boldsymbol{\lambda}})\, \mathsf{U}_1^{\dagger}\, \mathbf{f}_i\right]}{\mathbf{E}_{\mathcal{C}}\left[\max_i \mathbf{f}_i^{\dagger}\, \mathsf{U}_2 \,\mathsf{diag}(\underline{\boldsymbol{\mu}})\, \mathsf{U}_2^{\dagger}\, \mathbf{f}_i\right]} \geq \frac{\lambda_1}{\mu_1}. \tag{156}$$

From Lemma 1, we note that $\{\mathsf{U}_1^{\dagger}\, \mathbf{f}_i\}$ and $\{\mathsf{U}_2^{\dagger}\, \mathbf{f}_i\}$ are i.i.d. and have the same distribution as $\{\mathbf{f}_i\}$. Thus, (156) is equivalent to showing that

$$\frac{\mathbf{E}_{\mathcal{C}}\left[\max_i \mathbf{f}_i^{\dagger} \,\mathsf{diag}(\underline{\boldsymbol{\lambda}})\, \mathbf{f}_i\right]}{\lambda_1} \geq \frac{\mathbf{E}_{\mathcal{C}}\left[\max_i \mathbf{f}_i^{\dagger} \,\mathsf{diag}(\underline{\boldsymbol{\mu}})\, \mathbf{f}_i\right]}{\mu_1}. \tag{157}$$

In other words, the proof is complete if we can show that

$$f(\underline{\boldsymbol{\lambda}}) = \frac{\mathbf{E}_{\mathcal{C}}\left[\max_i \sum_k |\mathbf{f}_i(k)|^2 \lambda_k\right]}{\lambda_1} \tag{158}$$

is a Schur-concave function of $\underline{\boldsymbol{\lambda}}$.

It is important to note that $f(\underline{\boldsymbol{\lambda}})$ is a ratio of two Schur-convex functions. For this, it is obvious that $\underline{\boldsymbol{\lambda}} \prec \underline{\boldsymbol{\mu}}$ implies $\lambda_1 \leq \mu_1$. On the other hand, the numerator of $f(\underline{\boldsymbol{\lambda}})$ can be shown to be Schur-convex since $\max(\cdot)$ is a convex function of its argument. Without a standard recipe for studying the Schur-concavity of a ratio of Schur-convex functions, we resort to basic theory [42, A.2.b, p. 55] from which we can claim that $f(\cdot)$ is Schur-concave if and only if:

- $f(\cdot)$ is symmetric in its indices. That is, $f(\underline{\boldsymbol{\lambda}}) = f(\underline{\boldsymbol{\lambda}}\Pi)$ for all permutations $\Pi = [\pi_1, \cdots, \pi_{N_t}]$.
- $f\big([\lambda_1, s-\lambda_1, \lambda_3, \cdots, \lambda_{N_t}]\big)$ is decreasing in $\lambda_1$ for all $\lambda_1 \geq s/2$ and any choice of $s, \lambda_3, \cdots, \lambda_{N_t}$.

The first condition is straight-forward since

$$f(\underline{\boldsymbol{\lambda}}) = \frac{\mathbf{E}_{\mathcal{C}}\big[\max_i \sum_k |\mathbf{f}_i(k)|^2 \lambda_k\big]}{\max_k \lambda_k} \tag{159}$$

$$\stackrel{(a)}{=} \frac{\mathbf{E}_{\mathcal{C}}\big[\max_i \sum_k |\mathbf{f}_i(\pi_k)|^2 \lambda_{\pi_k}\big]}{\max_k \lambda_{\pi_k}} \tag{160}$$

$$\stackrel{(b)}{=} \frac{\mathbf{E}_{\mathcal{C}}\big[\max_i \sum_k |\mathbf{f}_i(k)|^2 \lambda_{\pi_k}\big]}{\max_k \lambda_{\pi_k}} = f(\underline{\boldsymbol{\lambda}}\Pi) \tag{161}$$

where (a) follows from the symmetricity of the sum function and (b) from the exchangeability of $|\mathbf{f}_i(k)|^2$ proved in Lemma 1. For the second condition, it can be seen that

$$f\Big([\lambda_1, s-\lambda_1, \lambda_3, \cdots, \lambda_{N_t}]\Big) = \mathbf{E}_{\mathcal{C}}\left[\max_i E_i\right] \tag{162}$$

$$E_i = |\mathbf{f}_i(1)|^2 - |\mathbf{f}_i(2)|^2 + \frac{\sum_{k\geq 2} |\mathbf{f}_i(k)|^2 \delta_k}{\lambda_1} \tag{163}$$

where $\delta_2 = s$, $\delta_k = \lambda_k$, $k \geq 3$. For every realization of $\{\mathbf{f}_i\}$ from the RVQ codebook and every choice of $s, \lambda_3, \cdots, \lambda_{N_t}$, all the functions $E_i$, $i = 1, \cdots, 2^B$ are decreasing in $\lambda_1$. Thus, the $\max(\cdot)$ function is also decreasing in $\lambda_1$. Averaging over the RVQ codebook, we arrive at the second condition. ■

### F. Proof of Theorem 3

In the $N_t = 2$ case, $\delta \triangleq \Delta_2 \cdot \log_e(2)$ is written as

$$\delta = \frac{\rho}{(\lambda_1-\lambda_2)^m} \int_{\lambda_2}^{\lambda_1} \frac{(x-\lambda_2)^m \, dx}{1+\rho x} \tag{164}$$

$$= \frac{\rho}{(\lambda_1-\lambda_2)^m} \int_0^{\lambda_1-\lambda_2} \frac{x^m dx}{1+\rho\lambda_2+\rho x} \tag{165}$$

$$= \frac{\rho}{(\lambda_1-\lambda_2)^m} \int_0^{\lambda_1-\lambda_2} \Big[\sum_{t=0}^{m-1} (-s)^t x^{m-1-t} + \frac{s^m}{x+s}\Big] dx \tag{166}$$

$$= \rho \left[\sum_{t=0}^{m-1} \left(\frac{-s}{\lambda_1-\lambda_2}\right)^t \cdot \frac{1}{m-t} + \left(\frac{s}{\lambda_1-\lambda_2}\right)^m \log_e\left(\frac{1+\rho\lambda_1}{1+\rho\lambda_2}\right)\right] \tag{167}$$

$$= \left(\frac{s}{\lambda_1-\lambda_2}\right)^m \left[\log_e(1+z) - \sum_{t=1}^{m} \frac{(-1)^{t+1} z^t}{t}\right] \tag{168}$$

where

$$s = \frac{1+\rho\lambda_2}{\rho}, \quad z = \frac{\lambda_1-\lambda_2}{s} = \frac{\rho(\lambda_1-\lambda_2)}{1+\rho\lambda_2}. \tag{169}$$

In the general $N_t$ case, the dominant term of $\delta = \Delta_2 \cdot \log_e(2)$ is written as

$$\delta \approx \int_0^{\lambda_1-\lambda_2} \left(1 - \left(\frac{y}{A}\right)^{N_t-1}\right)^m \cdot \frac{\rho dy}{1+\rho\lambda_1 - \rho y} \tag{170}$$

$$= \int_0^{\frac{\lambda_1-\lambda_2}{A}} \frac{\rho A \left(1 - y^{N_t-1}\right)^m dy}{1+\rho\lambda_1 - \rho A y} \tag{171}$$

where $A = \left(\prod_{j\geq 2} \lambda_1 - \lambda_j\right)^{\frac{1}{N_t-1}}$. There are two ways in which (171) can be computed: 1) replacing the denominator of the integrand by an appropriate geometric series, and 2) expanding the numerator of the integrand using the binomial theorem.

***Method 1:*** With $\gamma = \frac{\rho A}{1+\rho\lambda_1}$ and using the fact that $\gamma y < 1$ for all $y$ in (171), we replace the denominator in (171) with a geometric series to result in

$$\delta = \frac{\rho A}{1+\rho\lambda_1} \cdot \int_0^{\frac{\lambda_1-\lambda_2}{A}} \left(1 - y^{N_t-1}\right)^m \cdot \sum_{i=0}^{\infty} (\gamma y)^i. \tag{172}$$

Upon elementary integrand transformations, (172) is written as

$$\delta = \frac{2\rho A}{(N_t-1)(1+\rho\lambda_1)} \cdot \sum_{i=0}^{\infty} \gamma^i \int_0^{\theta_{\max}} \cos^{2m+1}(\theta) \sin^{p_i}(\theta) d\theta \tag{173}$$

where $\theta_{\max}$ is as in (139) and $p_i = \frac{2(i+1)}{N_t-1} - 1$. Computing this integral in closed-form using [50, 2.511(4), p. 131], we have

$$\delta = \frac{\rho A}{(N_t-1)(1+\rho\lambda_1)} \cdot \sum_{i=0}^{\infty} \frac{\gamma^i (1-D)^{\frac{i+1}{N_t-1}}}{m + \frac{i+1}{N_t-1}} \times \Bigg[ D^m + \sum_{k=1}^{m} \frac{2^k \cdot m(m-1)\cdots(m-k+1)}{(2m+p_i-1)\cdots(2m+p_i-2k+1)} D^{m-k} \Bigg]. \tag{174}$$

***Method 2:*** Alternately, expanding the numerator in (171) using the binomial theorem, we have

$$\delta = \frac{\rho A}{1+\rho\lambda_1} \int_0^{\frac{\lambda_1-\lambda_2}{A}} \frac{\sum_{k=0}^{m} \binom{m}{k} (-1)^k y^{(N_t-1)k} dy}{1-\gamma y} \tag{175}$$

$$= \frac{\rho A}{1+\rho\lambda_1} \cdot \sum_{k=0}^{m} \binom{m}{k} (-1)^k \frac{\left(\frac{\lambda_1-\lambda_2}{A}\right)^{(N_t-1)k+1}}{(N_t-1)k+1} \times {}_2F_1\left(1, (N_t-1)k+1; (N_t-1)k+2, \frac{\rho(\lambda_1-\lambda_2)}{1+\rho\lambda_1}\right) \tag{176}$$

where the second equation follows from [50, 3.194(5), p. 285], and

$${}_2F_1(a,b;c,z) = \sum_{n=0}^{\infty} \frac{(a)_n (b)_n}{(c)_n} \cdot \frac{z^n}{n!} \tag{177}$$

is the Gauss hypergeometric function with $(a)_n$ denoting the Pochhammer symbol:

$$(a)_n = a \cdot (a+1) \cdot \cdots \cdot (a+n-1),\ n \geq 1,\ (a)_0 = 1. \tag{178}$$

Using the definition of the hypergeometric function [50, 9.100, p. 1039], we have

$$\delta = \frac{\rho A}{1+\rho\lambda_1} \cdot \sum_{k=0}^{m} \binom{m}{k} (-1)^k \left(\frac{\lambda_1-\lambda_2}{A}\right)^{(N_t-1)k+1} \times$$
$$\sum_{i=0}^{\infty} \frac{1}{(N_t-1)k+1+i} \cdot \left(\frac{\rho(\lambda_1-\lambda_2)}{1+\rho\lambda_1}\right)^i. \quad (179)$$

The second expansion suffers from numerical instabilities due to the oscillatory nature (changing signs) of terms in the expansion.

***Correction Term:*** The expression for the correction term $\epsilon_B'$ in (70)-(71) and its trend in (72) follows on exactly the same lines as the proof of Theorem 1. Thus the details are not provided here. ∎

### G. Proof of Prop. 3

In the $N_t = 2$ case, as $B$ increases, two possibilities arise depending on the relationship between $\rho, \lambda_1$ and $\lambda_2$. In the first case, if

$$z < 1 \Longleftrightarrow \rho(\lambda_1 - 2\lambda_2) < 1, \quad (180)$$

using a Taylor's series approximation for $\log_e(1+z)$, we have

$$\delta \overset{B\to\infty}{\asymp} \left(\frac{s}{\lambda_1-\lambda_2}\right)^m \cdot \frac{z^{m+1}}{m+1} = \frac{z}{m+1}. \quad (181)$$

On the other hand, if

$$z \geq 1 \Longleftrightarrow \rho(\lambda_1 - 2\lambda_2) \geq 1, \quad (182)$$

using the fact that

$$\log_e(1+z) = \log_e(z) + \log_e\left(1+\frac{1}{z}\right), \quad (183)$$

we have

$$\delta \overset{B\to\infty}{\asymp} \frac{\log_e(z)+\frac{1}{z}}{z^m} + \sum_{t=0}^{m-1} \frac{(-1)^t}{z^t(m-t)} \quad (184)$$
$$\leq \frac{\log_e(z)+\frac{1}{z}}{z^m} + \left(1-\frac{1}{z}\right) \sum_{t=0}^{\frac{m}{2}-1} \frac{1}{z^{2t}(m-2t-1)} \quad (185)$$

where the second equation follows from the following reasoning:

$$\frac{1}{m-2t} < \frac{1}{m-2t-1},\ t = 0, \cdots, \frac{m}{2}-1. \quad (186)$$

We approximate the sum in (185) as $B \to \infty$ by the following integral:

$$\delta \overset{B\to\infty}{\asymp} \frac{\log_e(z)+\frac{1}{z}}{z^m} + \left(\frac{z-1}{2z}\right) \int_0^{\frac{m}{2}} \frac{e^{-\alpha t}}{\frac{m-1}{2}-t}\, dt \quad (187)$$

with $\alpha = 2\log_e(z) \geq 0$. Estimating the above integral from [50, 3.252(5-6), p. 311], we have

$$\delta \overset{B\to\infty}{\asymp} \frac{\log_e(z)+\frac{1}{z}}{z^m} + \left(\frac{z-1}{2z}\right)\cdot e^{-(m-1)\log_e(z)}\cdot\Big[\mathsf{Ei}\big((m-1)\log_e(z)\big) - \mathsf{Ei}\big(-\log_e(z)\big)\Big] \tag{188}$$

$$= \frac{\log_e(z)+\frac{1}{z}}{z^m} + \left(\frac{z-1}{2\cdot z^m}\right)\cdot\Big[\mathsf{E}_1\big(\log_e(z)\big) + \mathsf{li}\big(z^{m-1}\big)\Big] \tag{189}$$

where $\mathsf{E}_1(x) = \int_x^\infty \frac{e^{-t}dt}{t}$ and $\mathsf{Ei}(x) = -\mathsf{E}_1(-x)$ denote the exponential integral functions, and

$$\mathsf{li}(x) = \int_0^x \frac{dt}{\log_e(t)} \tag{190}$$

denotes the logarithmic integral function, respectively. From [51, p. 231], we have

$$\mathsf{li}(x) \overset{x\to\infty}{\asymp} \frac{x}{\log_e(x)} \Longrightarrow \delta \overset{B\to\infty}{\asymp} \frac{(z-1)}{2\cdot z\log_e(z)\cdot(m-1)}. \tag{191}$$

In the general $N_t$ case, it is easier to capture the asymptotic trends of $\Delta_2$ using the expression obtained from Method 1. For this, we first write

$$\frac{2^k\cdot m(m-1)\cdots(m-k+1)}{\prod_{j=1}^k(2m-2j+p_i+1)} = \frac{m!\cdot\Gamma(m-k+\frac{i+1}{N_t-1})}{\Gamma(m+\frac{i+1}{N_t-1})\cdot m-k!}. \tag{192}$$

Ignoring the term corresponding to $D^m$ in the inner sum in (174), $\delta$ can be rewritten as

$$\delta = \frac{1}{N_t-1}\sum_{i=1}^\infty e^{-\mu i}\left[\sum_{k=0}^{m-1} D^k \frac{m!\,\Gamma(k+\frac{i}{N_t-1})}{k!\,\Gamma(m+1+\frac{i}{N_t-1})}\right] \tag{193}$$

where $\mu = \log\left(\frac{1+\rho\lambda_1}{\rho(\lambda_1-\lambda_2)}\right) > 0$. We split the outer sum into two parts: $1 \leq i \leq N_t-1$ and $i \geq N_t$, and the inner sum into two parts: $k=0$ and $k \geq 1$ and denote the corresponding contributions to $\delta$ by $\delta_i$, $i = 1, \cdots, 4$ respectively.

With respect to $\delta_1$, we have

$$\delta_1 \triangleq \frac{1}{N_t-1}\sum_{i=1}^{N_t-1} e^{-\mu i}\cdot\frac{m!\cdot\Gamma\big(\frac{i}{N_t-1}\big)}{\Gamma(m+1+\frac{i}{N_t-1})} \tag{194}$$

$$\asymp \frac{\kappa}{N_t-1}\sum_{i=1}^{N_t-1} e^{-\mu i} m^{-\frac{i}{N_t-1}} \tag{195}$$

$$= \frac{\kappa}{N_t-1}\cdot\frac{1-\frac{e^{-\mu(N_t-1)}}{m}}{e^{\mu}m^{\frac{1}{N_t-1}}-1} \tag{196}$$

where the second line follows from the fact that $\Gamma(x)$ is monotonically decreasing in $0 < x \leq 1$ [51] and using the Stirling's formula for $\Gamma(\cdot)$. For $\delta_2$, we have

$$\delta_2 \triangleq \frac{1}{N_t - 1} \sum_{i=N_t}^{\infty} e^{-\mu i} \cdot \frac{\Gamma(m+1)\Gamma\left(\frac{i}{N_t-1}\right)}{\Gamma(m+1+\frac{i}{N_t-1})} \tag{197}$$

$$= \frac{1}{N_t - 1} \sum_{i=N_t}^{\infty} e^{-\mu i} \cdot \beta\left(m+1, \frac{i}{N_t-1}\right) \tag{198}$$

$$\leq \frac{1}{N_t - 1} \sum_{i=N_t}^{\infty} e^{-\mu i} \cdot \frac{1}{m+1} \cdot \frac{N_t - 1}{i} \tag{199}$$

$$\leq \frac{1}{m+1} \cdot \frac{1}{N_t - 1} \cdot \frac{e^{-\mu(N_t-1)}}{1 - e^{-\mu}} \tag{200}$$

where the second line follows from the definition of the Beta function, and the third line follows from the fact [52] that $\beta(x, y) \leq \frac{1}{xy}$ if $x \geq 1$ and $y \geq 1$. We now use the fact [53] that

$$\frac{\Gamma(n+1)}{\Gamma(n+s)} \geq n^{1-s}, \ 0 < s \leq 1 \tag{201}$$

to bound $\delta_3$ as follows:

$$\delta_3 \triangleq \frac{1}{N_t - 1} \sum_{i=1}^{N_t-1} \frac{e^{-\mu i} \cdot \Gamma(m+1)}{\Gamma\left(m+1+\frac{i}{N_t-1}\right)} \sum_{k=1}^{m-1} D^k \cdot \frac{\Gamma\left(k + \frac{i}{N_t-1}\right)}{\Gamma(k+1)} \tag{202}$$

$$\leq \frac{1}{N_t - 1} \sum_{i=1}^{N_t-1} e^{-\mu i} \frac{\Gamma(m+1)}{\Gamma\left(m+1+\frac{i}{N_t-1}\right)} \sum_{k=1}^{m-1} D^k \cdot k^{\frac{i}{N_t-1}-1} \tag{203}$$

$$\asymp \frac{D}{1-D} \cdot \frac{1}{N_t - 1} \sum_{i=1}^{N_t-1} e^{-\mu i} m^{-\frac{i}{N_t-1}} \tag{204}$$

$$= \frac{D}{1-D} \cdot \frac{1}{N_t - 1} \cdot \frac{1 - \frac{e^{-\mu(N_t-1)}}{m}}{e^{\mu} m^{\frac{1}{N_t-1}} - 1} \tag{205}$$

where the third line follows from Stirling's formula for $\Gamma(\cdot)$ and the fact that $k^{\frac{i}{N_t-1}-1}$ is a decreasing function of $k$.

For $\delta_4$, we have

$$\delta_4 \triangleq \frac{1}{N_t - 1} \sum_{i=N_t}^{\infty} e^{-\mu i} \sum_{k=1}^{m-1} D^k \frac{m!\,\Gamma(k+y)}{k!\,\Gamma(m+1+y)} \tag{206}$$

$$\leq \frac{1}{N_t - 1} \sum_{i=N_t}^{\infty} \frac{e^{-\mu i} \cdot m! \cdot e^{1-y}}{\Gamma(m+y+1)} \sum_{k=1}^{m-1} D^k \cdot \frac{(k+y)^{k+y-\frac{1}{2}}}{(k+1)^{k+\frac{1}{2}}} \tag{207}$$

where we use $y$ in (206) to denote $y = \frac{i}{N_t-1} \geq 1$ and the second line follows from [54], where if $b > a \geq 1$, we have

$$\frac{\Gamma(b)}{\Gamma(a)} < \frac{b^{b-\frac{1}{2}}}{a^{a-\frac{1}{2}}} \cdot e^{a-b}. \tag{208}$$

Using the fact that $\frac{(k+y)^{k+y-\frac{1}{2}}}{(k+1)^{k+\frac{1}{2}}}$ is monotonically increasing in $k$ for any $y \geq 1$, we have

$$\delta_4 \leq \frac{1}{N_t - 1} \sum_{i=N_t}^{\infty} \frac{e^{-\mu i} \cdot m! \cdot e^{1-y}}{\Gamma(m+y+1)} \sum_{k=1}^{m-1} D^k \cdot \frac{(m+y)^{m+y-\frac{1}{2}}}{(m+1)^{m+\frac{1}{2}}} \tag{209}$$

$$\leq \mathsf{C}_3 \cdot \sum_{i=N_t}^{\infty} \frac{e^{-\mu i} \cdot m! \cdot e^{1-y}}{\Gamma(m+y+1)} \cdot \frac{(m+y)^{m+y-\frac{1}{2}}}{(m+1)^{m+\frac{1}{2}}} \tag{210}$$

$$\asymp \mathsf{C}_3 \cdot \sum_{i=N_t}^{\infty} \frac{e^{-\mu i} \cdot (m+1)^{m+\frac{1}{2}} \cdot e}{(m+y+1)^{m+y+\frac{1}{2}}} \cdot \frac{(m+y)^{m+y-\frac{1}{2}}}{(m+1)^{m+\frac{1}{2}}} \tag{211}$$

$$= \mathsf{C}_3 \cdot \sum_{i=N_t}^{\infty} \frac{e^{-\mu i} \cdot e}{m+y+1} \cdot \left( \frac{m+y}{m+y+1} \right)^{m+y-\frac{1}{2}} \tag{212}$$

$$\asymp \mathsf{C}_3 \cdot \sum_{i=N_t}^{\infty} \frac{e^{-\mu i} \cdot e}{m+y+1} \cdot \exp\left( -\frac{m+y-\frac{1}{2}}{m+y+1} \right) \tag{213}$$

$$\leq \frac{\mathsf{C}_3}{m+1} \cdot \sum_{i=N_t}^{\infty} e^{-\mu i} = \frac{\mathsf{C}_3}{m+1} \cdot \frac{e^{-\mu(N_t-1)}}{1-e^{-\mu}} \tag{214}$$

where $\mathsf{C}_3 = \frac{D}{(1-D)(N_t-1)}$, the third line follows by using Stirling's formula for $\Gamma(m+1)$ (as a function of $m+1$) and $\Gamma(m+y+1)$ (as a function of $m+y$) and the fifth line follows from the fact that

$$(1+x)^{\frac{1}{x}} \overset{x \to 0}{\asymp} e. \tag{215}$$

Putting together the trends of $\delta_i$, $i = 1, \cdots, 4$, we obtain the conclusion in the statement of the proposition.

With respect to Method 2, we approximate the inner sum in (179) by an appropriate reformulation of the exponential integral, and as $B \to \infty$, we have

$$\delta \asymp \frac{\rho A}{1+\rho\lambda_1} \cdot \sum_{k=0}^{m} \binom{m}{k} (-1)^k \left( \frac{\lambda_1 - \lambda_2}{A} \right)^{(N_t-1)k+1} \times e^{\left((N_t-1)k+1\right)\mu} \cdot \mathsf{E}_1\Big( \big((N_t-1)k+1\big)\mu \Big) \tag{216}$$

$$= \sum_{k=0}^{m} \binom{m}{k} (-1)^k \left( \frac{1+\rho\lambda_1}{\rho A} \right)^{(N_t-1)k} \cdot \mathsf{E}_1\Big( \big((N_t-1)k+1\big)\mu \Big) \tag{217}$$

where $\mu = \log\left( \frac{1+\rho\lambda_1}{\rho(\lambda_1-\lambda_2)} \right) > 0$. The oscillatory nature (changing signs) of the terms in (217) and the intractable nature of the exponential integral (for general values of the argument) imply that it is much harder to obtain insights on the asymptotic trends of $\Delta_2$ with (217) than with the expression from Method 1. ■

*H. Proof of Prop. 4*

We can rewrite the distribution function relevant in computing $\Delta_{1,\mathsf{sk}}$ as follows:

$$\Pr\left(\frac{\mathbf{f}^\dagger \mathsf{A}^\dagger \mathsf{H}^\dagger \mathsf{H} \mathsf{A} \mathbf{f}}{\mathbf{f}^\dagger \mathsf{A}^\dagger \mathsf{A} \mathbf{f}} \leq x \Big| \mathbf{f}^\dagger \mathbf{f} = 1\right) = 1 - \Pr\Big(\mathbf{f}^\dagger \mathsf{A}^\dagger \left(\mathsf{H}^\dagger \mathsf{H} - x\mathbf{I}\right) \mathsf{A}\mathbf{f} \geq 0 \Big| \mathbf{f}^\dagger \mathbf{f} = 1\Big) \tag{218}$$

$$= 1 - \Pr\Big(\mathbf{f}^\dagger \mathsf{B}_x \mathbf{f} \geq 0 \Big| \mathbf{f}^\dagger \mathbf{f} = 1\Big) \tag{219}$$

where $\mathsf{B}_x$ is defined as

$$\mathsf{B}_x \triangleq \mathsf{A}^\dagger \mathsf{H}^\dagger \mathsf{H} \mathsf{A} - x \mathsf{A}^\dagger \mathsf{A}. \tag{220}$$

***Remark 1:*** Note that $\mathsf{B}_x$ is Hermitian, but not positive semi-definite. In fact, $\mathsf{B}_x$ has the same number of positive, negative, and zero eigenvalues as $\left(\mathsf{H}^\dagger \mathsf{H} - x\mathbf{I}\right) \mathsf{A}\mathsf{A}^\dagger$, which is the same as those of $\mathsf{H}^\dagger \mathsf{H} - x\mathbf{I}$ (see [55, Theorem 7.6.3, p. 465] for details). Using an eigen-decomposition for $\mathsf{B}_x$ of the form $\mathsf{B}_x = \mathsf{V}_x \mathsf{\Gamma}_x \mathsf{V}_x^\dagger$ in the special case of $N_t = 2$ where $\mathsf{\Gamma}_x = \mathsf{diag}\big([\Gamma_{1,x},\, \Gamma_{2,x}]\big)$ such that $\Gamma_{1,x} \geq \Gamma_{2,x}$, we have:

1) $\Gamma_{1,x} \geq 0 = \Gamma_{2,x}$ if $x = \lambda_2(\mathsf{H}^\dagger \mathsf{H})$,
2) $\Gamma_{1,x} \geq 0 \geq \Gamma_{2,x}$ if $x \in (\lambda_2(\mathsf{H}^\dagger \mathsf{H}),\, \lambda_1(\mathsf{H}^\dagger \mathsf{H}))$,
3) $\Gamma_{1,x} = 0 \geq \Gamma_{2,x}$ if $x = \lambda_1(\mathsf{H}^\dagger \mathsf{H})$.

■

Thus, we can rewrite (219) as

$$\Pr\left(\frac{\mathbf{f}^\dagger \mathsf{A}^\dagger \mathsf{H}^\dagger \mathsf{H} \mathsf{A} \mathbf{f}}{\mathbf{f}^\dagger \mathsf{A}^\dagger \mathsf{A} \mathbf{f}} \leq x \Big| \mathbf{f}^\dagger \mathbf{f} = 1\right) = 1 - \Pr\Big(|\mathbf{f}(1)|^2 \Gamma_{1,x} + |\mathbf{f}(2)|^2 \Gamma_{2,x} \geq 0 \Big| \mathbf{f}^\dagger \mathbf{f} = 1\Big) \tag{221}$$

$$= 1 - \Pr\Big(|\mathbf{f}(1)|^2 \left(\Gamma_{1,x} - \Gamma_{2,x}\right) \geq -\Gamma_{2,x} \Big| \mathbf{f}^\dagger \mathbf{f} = 1\Big) \tag{222}$$

$$= 1 - \Pr\left(|\mathbf{f}(1)|^2 \geq \frac{|\Gamma_{2,x}|}{|\Gamma_{1,x}| + |\Gamma_{2,x}|} \Big| \mathbf{f}^\dagger \mathbf{f} = 1\right) \tag{223}$$

where the second equation follows from noting that $\Gamma_{1,x} \geq 0$ and $\Gamma_{2,x} \leq 0$ for all $x \in [\lambda_2(\mathsf{H}^\dagger \mathsf{H}),\, \lambda_1(\mathsf{H}^\dagger \mathsf{H})]$. We now use [8, Lemmas 2 and 4] to compute the above term (see Appendix B for details) as

$$\Pr\left(|\mathbf{f}(1)|^2 \geq \frac{|\Gamma_{2,x}|}{|\Gamma_{1,x}| + |\Gamma_{2,x}|} \Big| \mathbf{f}^\dagger \mathbf{f} = 1\right) = \frac{|\Gamma_{1,x}|}{|\Gamma_{1,x}| + |\Gamma_{2,x}|}. \tag{224}$$

Thus, $\Delta_{1,\mathsf{sk}}$ can be expressed as

$$\Delta_{1,\mathsf{sk}} = \frac{1}{\lambda_1(\mathsf{H}^\dagger \mathsf{H})} \cdot \int_{\lambda_2(\mathsf{H}^\dagger \mathsf{H})}^{\lambda_1(\mathsf{H}^\dagger \mathsf{H})} \left(\frac{|\Gamma_{2,x}|}{|\Gamma_{1,x}| + |\Gamma_{2,x}|}\right)^m dx. \tag{225}$$

Now observe that $\Delta_{1,\mathsf{sk}}$ is monotonically increasing as a function of $\frac{|\Gamma_{2,x}|}{|\Gamma_{1,x}|}$. Thus, an upper bound on $\frac{|\Gamma_{2,x}|}{|\Gamma_{1,x}|}$ also results in a corresponding upper bound on $\Delta_{1,\mathsf{sk}}$. For this, note that

$$\Gamma_{2,x} = \lambda_2(\mathsf{A}^\dagger \mathsf{H}^\dagger \mathsf{H} \mathsf{A} - x \mathsf{A}^\dagger \mathsf{A}) \tag{226}$$

$$\geq \lambda_2(\mathsf{A}^\dagger \mathsf{H}^\dagger \mathsf{H} \mathsf{A}) - x \lambda_1(\mathsf{A}^\dagger \mathsf{A}) \tag{227}$$

where the second step follows from a routine application of Weyl's inequality [55]. Since the right-hand side of (227) is non-positive for all $x$, we thus have

$$|\Gamma_{2,x}| \leq x\lambda_1(\mathsf{A}^\dagger\mathsf{A}) - \lambda_2(\mathsf{A}^\dagger\mathsf{H}^\dagger\mathsf{H}\mathsf{A}). \tag{228}$$

For $\Gamma_{1,x}$, we use [56, Corollary 11] to see that

$$\Gamma_{1,x} \geq \big(\lambda_1(\mathsf{H}^\dagger\mathsf{H}) - x\big) \cdot \lambda_1(\mathsf{A}\mathsf{A}^\dagger). \tag{229}$$

Note that the bounds in (228) and (229) are non-trivial (that is, the bounding terms are non-negative). Combining them, we have

$$\frac{|\Gamma_{2,x}|}{|\Gamma_{1,x}|} \leq \frac{x\lambda_1(\mathsf{A}^\dagger\mathsf{A}) - \lambda_2(\mathsf{A}^\dagger\mathsf{H}^\dagger\mathsf{H}\mathsf{A})}{(\lambda_1(\mathsf{H}^\dagger\mathsf{H}) - x) \cdot \lambda_1(\mathsf{A}\mathsf{A}^\dagger)}. \tag{230}$$

Using the bound in (230), after a routine integral computation, it is straightforward to see that

$$\Delta_{1,\,\mathsf{sk}} \leq \underbrace{\frac{1}{m+1} \cdot \left[1 - \frac{\lambda_2(\mathsf{A}^\dagger\mathsf{H}^\dagger\mathsf{H}\mathsf{A})}{\lambda_1(\mathsf{H}^\dagger\mathsf{H}) \cdot \lambda_1(\mathsf{A}\mathsf{A}^\dagger)} \cdot (1 - (\mathsf{C}_4)^m) - \frac{\lambda_2(\mathsf{H}^\dagger\mathsf{H})}{\lambda_1(\mathsf{H}^\dagger\mathsf{H})} \cdot (\mathsf{C}_4)^m\right]}_{\overline{\Delta}_{1,\,\mathsf{sk}}} \tag{231}$$

where $m = 2^B$ and

$$\mathsf{C}_4 = \frac{\lambda_2(\mathsf{H}^\dagger\mathsf{H}) \cdot \lambda_1(\mathsf{A}\mathsf{A}^\dagger) - \lambda_2(\mathsf{A}^\dagger\mathsf{H}^\dagger\mathsf{H}\mathsf{A})}{\lambda_1(\mathsf{H}^\dagger\mathsf{H}) \cdot \lambda_1(\mathsf{A}\mathsf{A}^\dagger) - \lambda_2(\mathsf{A}^\dagger\mathsf{H}^\dagger\mathsf{H}\mathsf{A})}. \tag{232}$$

Since $\mathsf{C}_4 \leq 1$ and $B \to \infty$, the conclusion in (81) is immediate. ■

### *I. Proof of Theorem 4*

For any integer $k \geq 1$, define the following expectation over the ensemble of RVQ codebooks

$$G_k \triangleq \left(\mathbf{E}_{\mathcal{C}}\left[\left(\frac{\mathbf{f}^\dagger\mathsf{A}^\dagger\mathsf{H}^\dagger\mathsf{H}\mathsf{A}\mathbf{f}}{\mathbf{f}^\dagger\mathsf{A}^\dagger\mathsf{A}\mathbf{f}}\right)^k\right]\right)^{\frac{1}{k}}. \tag{233}$$

It is easy to check that

$$\lim_{k\to\infty} G_k = \lambda_1(\mathsf{H}^\dagger\mathsf{H}) \tag{234}$$

$$G_1 \geq \lambda_{N_t}(\mathsf{H}^\dagger\mathsf{H}). \tag{235}$$

and it follows from Lyapunov's inequality [57, Prob. 28, p. 143] that $G_k$ is non-decreasing with $k$. Since

$$\lambda_{N_t}(\mathsf{H}^\dagger\mathsf{H}) \leq \frac{\lambda_1(\mathsf{A}^\dagger\mathsf{H}^\dagger\mathsf{H}\mathsf{A})}{\lambda_1(\mathsf{A}^\dagger\mathsf{A})} \leq \lambda_1(\mathsf{H}^\dagger\mathsf{H}), \tag{236}$$

it can be concluded that there exists some $K_\mathsf{L} \geq 1$ and some $K_\mathsf{U}$ satisfying $K_\mathsf{L} \leq K_\mathsf{U} < \infty$ such that

$$G_{K_\mathsf{L}-1} \leq \frac{\lambda_1(\mathsf{A}^\dagger\mathsf{H}^\dagger\mathsf{H}\mathsf{A})}{\lambda_1(\mathsf{A}^\dagger\mathsf{A})} \leq G_{K_\mathsf{L}} \tag{237}$$

$$\lambda_1(\mathsf{H}^\dagger\mathsf{H}) - 2^{-\frac{B}{2}} \leq G_{K_\mathsf{U}} \leq \lambda_1(\mathsf{H}^\dagger\mathsf{H}). \tag{238}$$

Thus, $\Delta_{1,\mathsf{sk}}$ can be bounded as

$$\Delta_{1,\mathsf{sk}} \cdot \lambda_1(\mathsf{H}^\dagger\mathsf{H}) \leq \int_{\lambda_{N_t}(\mathsf{H}^\dagger\mathsf{H})}^{\frac{\lambda_1(\mathsf{A}^\dagger\mathsf{H}^\dagger\mathsf{HA})}{\lambda_1(\mathsf{A}^\dagger\mathsf{A})}} \left[\Pr\left(\frac{\mathbf{f}^\dagger\mathsf{A}^\dagger\mathsf{H}^\dagger\mathsf{HA}\mathbf{f}}{\mathbf{f}^\dagger\mathsf{A}^\dagger\mathsf{A}\mathbf{f}} \leq x \Big| \mathbf{f}^\dagger\mathbf{f} = 1\right)\right]^m dx$$
$$+ \sum_{k=K_{\mathsf{L}}}^{K_{\mathsf{U}}} \int_{G_{k-1}}^{G_k} \left[\Pr\left(\frac{\mathbf{f}^\dagger\mathsf{A}^\dagger\mathsf{H}^\dagger\mathsf{HA}\mathbf{f}}{\mathbf{f}^\dagger\mathsf{A}^\dagger\mathsf{A}\mathbf{f}} \leq x \Big| \mathbf{f}^\dagger\mathbf{f} = 1\right)\right]^m dx$$
$$+ \int_{G_{K_{\mathsf{U}}}}^{\lambda_1(\mathsf{H}^\dagger\mathsf{H})} \left[\Pr\left(\frac{\mathbf{f}^\dagger\mathsf{A}^\dagger\mathsf{H}^\dagger\mathsf{HA}\mathbf{f}}{\mathbf{f}^\dagger\mathsf{A}^\dagger\mathsf{A}\mathbf{f}} \leq x \Big| \mathbf{f}^\dagger\mathbf{f} = 1\right)\right]^m dx \tag{239}$$
$$\triangleq \overline{\Delta}_{1,\mathsf{sk}}. \tag{240}$$

For the first term of $\overline{\Delta}_{1,\mathsf{sk}}$ in (239) (denoted as $\mathcal{T}_1$), since $\mathbf{f}^\dagger\mathsf{A}^\dagger\mathsf{A}\mathbf{f} \leq \lambda_1(\mathsf{A}^\dagger\mathsf{A})$, we have

$$\Pr\left(\frac{\mathbf{f}^\dagger\mathsf{A}^\dagger\mathsf{H}^\dagger\mathsf{HA}\mathbf{f}}{\mathbf{f}^\dagger\mathsf{A}^\dagger\mathsf{A}\mathbf{f}} \leq x \Big| \mathbf{f}^\dagger\mathbf{f} = 1\right) \leq \Pr\left(\frac{\mathbf{f}^\dagger\mathsf{A}^\dagger\mathsf{H}^\dagger\mathsf{HA}\mathbf{f}}{\lambda_1(\mathsf{A}^\dagger\mathsf{A})} \leq x \Big| \mathbf{f}^\dagger\mathbf{f} = 1\right) \tag{241}$$
$$= 1 - \frac{\left(\lambda_1(\mathsf{A}^\dagger\mathsf{H}^\dagger\mathsf{HA}) - x\lambda_1(\mathsf{A}^\dagger\mathsf{A})\right)^{N_t-1}}{\prod_{j=2}^{N_t} \lambda_1(\mathsf{A}^\dagger\mathsf{H}^\dagger\mathsf{HA}) - \lambda_j(\mathsf{A}^\dagger\mathsf{H}^\dagger\mathsf{HA})} \tag{242}$$

where the second step follows from an application of Lemma 2 to the distribution of $\mathbf{f}^\dagger\mathsf{A}^\dagger\mathsf{H}^\dagger\mathsf{HA}\mathbf{f}$. Using a computation that mirrors that in Theorem 1, we have

$$\mathcal{T}_1 \overset{B\to\infty}{\asymp} \frac{\kappa \cdot 2^{-\frac{B}{N_t-1}}}{N_t - 1} \cdot \left(\frac{\lambda_1(\mathsf{A}^\dagger\mathsf{H}^\dagger\mathsf{HA})}{\lambda_1(\mathsf{A}^\dagger\mathsf{A})} - \lambda_{N_t}(\mathsf{H}^\dagger\mathsf{H})\right) \cdot \left(1 + \frac{D_{\mathsf{sk}}}{(1 - D_{\mathsf{sk}})(N_t - 1)}\right) \tag{243}$$

with $\kappa = \Gamma\left(\frac{1}{N_t - 1}\right)$ and

$$D_{\mathsf{sk}} = 1 - \prod_{j=2}^{N_t} \frac{\lambda_1(\mathsf{A}^\dagger\mathsf{H}^\dagger\mathsf{HA}) - \lambda_{N_t}(\mathsf{H}^\dagger\mathsf{H}) \cdot \lambda_1(\mathsf{A}^\dagger\mathsf{A})}{\lambda_1(\mathsf{A}^\dagger\mathsf{H}^\dagger\mathsf{HA}) - \lambda_j(\mathsf{A}^\dagger\mathsf{H}^\dagger\mathsf{HA})}. \tag{244}$$

The tightness of (243) follows from the tightness result established in Theorem 1.

For bounding the second term of (239) (denoted as $\mathcal{T}_2$), we need a reverse Cauchy-Schwarz inequality, which is presented next.

*Lemma 5:* Let $\mathbf{X}$ be a positive random variable. Let $g(\cdot) : \mathbb{R}^+ \mapsto \mathbb{R}^+$ be a monotonically increasing function such that $g(\mathbf{X})$ and $(g(\mathbf{X}))^2$ are integrable. If $x$ is such that $g(x) \leq \mathbf{E}\left[g(\mathbf{X})\right]$, we have

$$\Pr\left(\mathbf{X} > x\right) \geq \frac{\Big(\mathbf{E}[g(\mathbf{X})] - g(x)\Big)^2}{\mathbf{E}\left[\big(g(\mathbf{X})\big)^2\right]}. \tag{245}$$

*Proof:* Since $x$ is such that $\mathbf{E}\left[g(\mathbf{X})\right] \geq g(x)$, using the standard Cauchy-Schwarz inequality and the monotonicity of $g(\cdot)$, we have

$$\mathbf{E}[g(\mathbf{X})] - g(x) \leq \mathbf{E}[g(\mathbf{X})] - \mathbf{E}[g(\mathbf{X})\mathbb{1}(\mathbf{X} \leq x)] \tag{246}$$

$$= \mathbf{E}\left[g(\mathbf{X}) - g(\mathbf{X})\mathbb{1}(\mathbf{X} \leq x)\right] \tag{247}$$

$$= \mathbf{E}\left[g(\mathbf{X})\mathbb{1}(\mathbf{X} > x)\right] \tag{248}$$

$$\leq \sqrt{\mathbf{E}\left[(g(\mathbf{X}))^2\right] \cdot \Pr(\mathbf{X} > x)}. \tag{249}$$

Rearranging (249), we have the conclusion of the lemma. ■

For each $k$ satisfying $K_{\mathsf{L}} \leq k \leq K_{\mathsf{U}}$, we repeatedly apply Lemma 5 with

$$\mathbf{X} = \frac{\mathbf{f}^\dagger \mathsf{A}^\dagger \mathsf{H}^\dagger \mathsf{H} \mathsf{A} \mathbf{f}}{\mathbf{f}^\dagger \mathsf{A}^\dagger \mathsf{A} \mathbf{f}} \tag{250}$$

and $g(x) = x^k$ to get the following bound for $\mathcal{T}_2$:

$$\mathcal{T}_2 \leq \sum_{k=K_{\mathsf{L}}}^{K_{\mathsf{U}}} \int_{G_{k-1}}^{G_k} \left[1 - \left(\frac{(G_k)^k - x^k}{(G_{2k})^k}\right)^2\right]^m dx \tag{251}$$

$$= \sum_{k=K_{\mathsf{L}}}^{K_{\mathsf{U}}} \frac{(G_{2k})^k}{k} \int_0^{I_k} \frac{(1-y^2)^m \, dy}{\left((G_k)^k - y(G_{2k})^k\right)^{\frac{k-1}{k}}} \tag{252}$$

$$\leq \sum_{k=K_{\mathsf{L}}}^{K_{\mathsf{U}}} \frac{(G_{2k})^k}{k \cdot (G_{k-1})^{k-1}} \int_0^{I_k} (1-y^2)^m \, dy \tag{253}$$

where $I_k = \frac{(G_k)^k - (G_{k-1})^k}{(G_{2k})^k}$, the second equation follows from a transformation $y \mapsto \frac{(G_k)^k - x^k}{(G_{2k})^k}$, and the third step follows by trivially bounding $y \leq I_k$. Note that the monotonicity of $G_k$ with $k$ implies that $I_k \leq 1$. With the transformation $y \mapsto \sin(\theta)$, we can reuse the computation in Theorem 1 to estimate $\mathcal{T}_2$. However, this estimate is not sufficient for our purpose and hence, we will establish a tighter estimate now.

From [50, 2.512(3), p. 131] and Stirling's formula for $h(m)$ in (146), we have

$$\mathcal{T}_2 \asymp \sum_{k=K_{\mathsf{L}}}^{K_{\mathsf{U}}} \frac{(G_{2k})^k}{k \cdot (G_{k-1})^{k-1}} \cdot \frac{\sqrt{\pi m} \cdot I_k}{2m+1} \cdot \left(1 + \sum_{j=1}^{m-1} \frac{(1-I_k^2)^j}{h(j)}\right). \tag{254}$$

Since $h(j) \asymp \sqrt{\pi j}$ for $j$ large, we can estimate (254) by

$$\mathcal{T}_2 \cdot (2m+1) \asymp \sum_{k=K_{\mathsf{L}}}^{K_{\mathsf{U}}} \frac{(G_{2k})^k \cdot \sqrt{\pi m} \cdot I_k}{k \cdot (G_{k-1})^{k-1}} \left(1 + \int_1^\infty \frac{e^{-\alpha_k x}}{\sqrt{\pi x}} dx\right) \tag{255}$$

$$= \sum_{k=K_{\mathsf{L}}}^{K_{\mathsf{U}}} \frac{(G_{2k})^k \cdot \sqrt{\pi m} \cdot I_k}{k \cdot (G_{k-1})^{k-1}} \left(1 + \frac{\Gamma\left(1/2, \alpha_k\right)}{\sqrt{\pi \alpha_k}}\right) \tag{256}$$

$$= \sum_{k=K_{\mathsf{L}}}^{K_{\mathsf{U}}} \frac{(G_{2k})^k \cdot \sqrt{\pi m} \cdot I_k}{k \cdot (G_{k-1})^{k-1}} \left(1 + \frac{1 - \mathsf{erf}(\sqrt{\alpha_k})}{\sqrt{\alpha_k}}\right) \tag{257}$$

where $\alpha_k = \log_e\left(\frac{1}{1-I_k^2}\right)$, and

$$\Gamma(a,x) = \int_x^{\infty} t^{a-1} e^{-t} dt \tag{258}$$

is the incomplete Gamma function. The second step follows from [51, 6.5.3, p. 260], and

$$\mathsf{erf}(x) = \frac{2}{\sqrt{\pi}} \int_0^x e^{-t^2} dt \tag{259}$$

is the error function. The third step follows from [51, 6.5.17, p. 262]. Note that as $B$ increases, $K_{\mathsf{U}}$ increases and $I_k \to 0$. As a result, we have

$$\alpha_k = \log_e\left(1 + \frac{I_k^2}{1-I_k^2}\right) \overset{B\to\infty}{\asymp} I_k^2. \tag{260}$$

In this setting, from [51, 7.1.6, p. 297], we thus have

$$\mathcal{T}_2 \overset{B\to\infty}{\asymp} 2^{-\frac{B}{2}} \sum_{k=K_{\mathsf{L}}}^{K_{\mathsf{U}}} \mathsf{C}_5 \cdot \frac{(G_{2k})^k}{k \cdot (G_{k-1})^{k-1}} \tag{261}$$

for some constant $\mathsf{C}_5$. Using the relationship in (237)-(238), we can write (261) as

$$\mathcal{T}_2 \overset{B\to\infty}{\asymp} \mathsf{C}_5 \cdot 2^{-\frac{B}{2}} \cdot \left( \frac{(G_{2K_{\mathsf{L}}})^{K_{\mathsf{L}}}}{K_{\mathsf{L}} \cdot (G_{K_{\mathsf{L}}-1})^{K_{\mathsf{L}}-1}} + \sum_{k=K_{\mathsf{L}}}^{K_{\mathsf{U}}-1} \frac{\lambda_1(\mathsf{H}^\dagger\mathsf{H})}{k+1} \left( \frac{\lambda_1(\mathsf{H}^\dagger\mathsf{H})\lambda_1(\mathsf{A}\mathsf{A}^\dagger)}{\lambda_1(\mathsf{A}^\dagger\mathsf{H}^\dagger\mathsf{H}\mathsf{A})} \right)^k \right) \tag{262}$$

$$\leq 2^{-\frac{B}{2}} \cdot \mathsf{C}_5 \cdot \lambda_1(\mathsf{H}^\dagger\mathsf{H}) \cdot \left( \frac{1}{K_{\mathsf{L}}} \cdot \left( \frac{\lambda_1(\mathsf{H}^\dagger\mathsf{H})}{\lambda_{N_t}(\mathsf{H}^\dagger\mathsf{H})} \right)^{K_{\mathsf{L}}-1} + \sum_{k=K_{\mathsf{L}}}^{K_{\mathsf{U}}-1} \frac{1}{k+1} \left( \frac{\lambda_1(\mathsf{H}^\dagger\mathsf{H})\lambda_1(\mathsf{A}\mathsf{A}^\dagger)}{\lambda_1(\mathsf{A}^\dagger\mathsf{H}^\dagger\mathsf{H}\mathsf{A})} \right)^k \right) \tag{263}$$

$$\triangleq 2^{-\frac{B}{2}} \cdot G\left( \frac{\lambda_1(\mathsf{H}^\dagger\mathsf{H}) \cdot \lambda_1(\mathsf{A}\mathsf{A}^\dagger)}{\lambda_1(\mathsf{A}^\dagger\mathsf{H}^\dagger\mathsf{H}\mathsf{A})} \right) \tag{264}$$

where we have used the symbolic notation $G(\cdot)$ to denote the monotonically increasing function in (263) for a given $\mathsf{H}$. The tightness of (264) is due to the tight estimation of the integral in (253).

For the third term of (239) (denoted as $\mathcal{T}_3$), we trivially over-bound $\Pr\left( \frac{\mathbf{f}^\dagger\mathsf{A}^\dagger\mathsf{H}^\dagger\mathsf{H}\mathsf{A}\mathbf{f}}{\mathbf{f}^\dagger\mathsf{A}^\dagger\mathsf{A}\mathbf{f}} \leq x \middle| \mathbf{f}^\dagger\mathbf{f} = 1 \right)$ by $1$ and use the definition of $K_{\mathsf{U}}$ to obtain

$$\mathcal{T}_3 \leq \lambda_1(\mathsf{H}^\dagger\mathsf{H}) - G_{K_{\mathsf{U}}} \leq 2^{-\frac{B}{2}}. \tag{265}$$

Combining the three terms $\mathcal{T}_1$, $\mathcal{T}_2$ and $\mathcal{T}_3$, we have

$$\Delta_{1,\mathsf{sk}} \cdot \lambda_1(\mathsf{H}^\dagger\mathsf{H}) \leq \overline{\Delta}_{1,\mathsf{sk}} \cdot \lambda_1(\mathsf{H}^\dagger\mathsf{H}) \tag{266}$$

$$\overset{B\to\infty}{\asymp} 2^{-\frac{B}{2}} \cdot \left( 1 + G\left( \frac{\lambda_1(\mathsf{H}^\dagger\mathsf{H})\lambda_1(\mathsf{A}\mathsf{A}^\dagger)}{\lambda_1(\mathsf{A}^\dagger\mathsf{H}^\dagger\mathsf{H}\mathsf{A})} \right) \right) + \frac{\kappa \cdot 2^{-\frac{B}{N_t-1}}}{N_t - 1} \times \left( \frac{\lambda_1(\mathsf{A}^\dagger\mathsf{H}^\dagger\mathsf{H}\mathsf{A})}{\lambda_1(\mathsf{A}^\dagger\mathsf{A})} - \lambda_{N_t}(\mathsf{H}^\dagger\mathsf{H}) \right) \cdot \left( 1 + \frac{D_{\mathsf{sk}}}{(1-D_{\mathsf{sk}})(N_t-1)} \right). \tag{267}$$

If $N_t \geq 4$, it is clear that the first term in (267) is sub-dominant relative to the second term. The statement of the theorem hence follows. ■